\documentclass[prb,aps,twocolumn,superscriptaddress]{revtex4-2}
\usepackage{slashed}
\usepackage{tikz-cd}
	\usepackage{amsmath}
	\usepackage{amsfonts}
	\usepackage{amssymb}
	\usepackage{bbm}
	\usepackage{graphicx}
	\usepackage{mathtools}
	\usepackage{stmaryrd}
	\usepackage[force]{feynmp-auto}
	\usepackage[colorlinks=true,citecolor=blue,linkcolor=red]{hyperref}
	\usepackage{bbold}					% for \mathbb{1} as a nice unit matrix symbol
	\usepackage[makeroom]{cancel}		% crossed terms in tables
	\usepackage{multirow}				% merge cells vertically in tables
	\usepackage[normalem]{ulem}        % strike-through text
	\usepackage{array}

	\newcolumntype{x}[1]{>{\centering\let\newline\\\arraybackslash\hspace{0pt}}p{#1}}
	\newcommand*{\Ang}{\ensuremath{{\mbox{\normalfont\AA}}}}

	\makeatletter
\newcommand*\rel@kern[1]{\kern#1\dimexpr\macc@kerna}
\newcommand*\widebar[1]{%
  \begingroup
  \def\mathaccent##1##2{%
    \rel@kern{0.8}%
    \overline{\rel@kern{-0.8}\macc@nucleus\rel@kern{0.2}}%
    \rel@kern{-0.2}%
  }%
  \macc@depth\@ne
  \let\math@bgroup\@empty \let\math@egroup\macc@set@skewchar
  \mathsurround\z@ \frozen@everymath{\mathgroup\macc@group\relax}%
  \macc@set@skewchar\relax
  \let\mathaccentV\macc@nested@a
  \macc@nested@a\relax111{#1}%
  \endgroup
}
\makeatother
	
\makeatletter
\newcommand*\wt[1]{\mathpalette\wthelper{#1}}
\newcommand*\wthelper[2]{%
    \hbox{\dimen@\accentfontxheight#1%
        \accentfontxheight#11.15\dimen@
        $\m@th#1\widetilde{#2}$%
        \accentfontxheight#1\dimen@
    }%
}
\newcommand*\accentfontxheight[1]{%
    \fontdimen5\ifx#1\displaystyle
        \textfont
    \else\ifx#1\textstyle
        \textfont
    \else\ifx#1\scriptstyle
        \scriptfont
    \else
        \scriptscriptfont
    \fi\fi\fi3
}
\makeatother

	\def\bk{{\bf{k}}}
	\def\br{{\bf{r}}}
	\def\bx{{\bf{x}}}
	
	\def\bu{{\bf{u}}}
	\def\bB{{\bf{B}}}
	\def\bPi{{\bf{\Pi}}}

\begin{document}

\author{Xiao-Qi~Sun}
\affiliation{Department of Physics and Institute for Condensed Matter Theory,
University of Illinois at Urbana-Champaign, Urbana, IL 61801, USA}
\author{Jing-Yuan~Chen}
\affiliation{Institute for Advanced Study, Tsinghua University, Beijing, 100084, China}
\author{Steven~A.~Kivelson}
\affiliation{Department of Physics,
Stanford University, Stanford, CA 93405, USA}

\title{Large extrinsic phonon thermal Hall effect from resonant scattering}
\date{\today}

\begin{abstract}
 Recent experimental observations of unexpectedly large thermal Hall conductivities, $\kappa_H$, in insulating materials, including the parent compounds of the high temperature superconducting cuprates, likely reflect an extrinsic contribution  from a yet to be identified extrinsic source of skew scattering of acoustic phonons. We show that resonant scattering of phonons from a certain class of three-level systems produces strong skew scattering in the presence of a modest magnetic field. We interpret this as a first step towards understanding the experiments.
\end{abstract}

\maketitle

\section{Introduction}
The contribution of charge neutral excitations to the thermal Hall conductance, $\kappa_H$, is generally expected to be small because they couple relatively weakly to an applied magnetic field.  However, since at low temperature $T$ the only thermally accessible excitations are acoustic phonons, recent experimental observations of unexpectedly large magnitudes of $\kappa_H$ in the para-electric insulator $\text{SrTiO}_3$~\cite{Li:2020} and a number of cuprate Mott insulators~\cite{Grissonnanche:2019,boulanger2020thermal,Grissonnanche:2020} suggest that, {\em at least} in these cases, the thermal Hall current is carried by phonons. As is the case with the electrical Hall effect, in the presence of a magnetic field a finite $\kappa_H$ is allowed by symmetry, even in the ideal limit of no scattering; this is known as the intrinsic effect~\cite{Qin:2012}.  However, there is also the possibility of an extrinsic effect which is proportional to the phonon mean free path;  at low $T$, where acoustic phonon mean free paths are often  large, it is likely that the extrinsic effect is always dominant.  

Treating phonon heat transport using the Boltzmann equation, and neglecting a typically unimportant intrinsic contribution from  the Berry curvature of the phonon wave-functions, it is straightforward to see that at low $T$
\begin{equation}
    \kappa_L \sim  C_v\, v^2 \, \tau, \ \ \ \ \
    \kappa_H  \sim C_v\, v^2 \, \tau^2 \tau_o^{-1} 
    \label{kappaL_kappaH}
\end{equation}
where in the familiar expression for $\kappa_L$, $C_v$ is the phonon contribution to the specific heat (i.e. neglecting contributions from nuclear spins etc.), $v$ is the speed of sound, and $\tau^{-1}$ is an appropriate average of the phonon scattering rates, while in the expression for $\kappa_H$ which we will derive later, $\tau_o^{-1}$ is the time-reversal odd \emph{skew scattering} rate to linear order in the magnetic field $B$. (Unless otherwise specified, we will henceforth use units such that $k_B=\hbar= 1$. ) 

It is common in crystalline insulators that the mean free path of thermal phonons at low $T$ is set by the sample size~\cite{Casimir:1938}.  In part, this reflects the fact that inelastic phonon-phonon scattering becomes ineffective at low $T$~\cite{Pomeranchuk:1941,Klemens:1951}.  However, even elastic scattering off defects becomes rapidly weak since low energy phonons are Goldstone modes. Nonetheless, it was proposed in Ref.~\cite{Chen:2020} that scattering of phonons off extended defects (tentatively identified as twin boundaries) can account for the experimentally observed temperature independent mean free path that determines $\kappa_L$  and the low temperature scaling and magnitude of 
$\kappa_H$ in  $\text{SrTiO}_3$~\cite{Li:2020}.

More generally, the origin of such scattering remains to be determined. Subsequent perturbative calculations~\cite{Guo:2021} have shown that scattering from \emph{non-dynamical} point defects is weak for long wavelength phonons and is unlikely to account for the large thermal Hall effect observed. In contrast, in the classic works~\cite{Anderson:1972,Phillips:1972} on the theory of insulating glasses, it was shown that resonant scattering of phonons off \emph{dynamical} two-level systems dominates the thermal transport. On the basis of these two observations, we are led to propose that some form of resonant skew scattering of phonons is an essential ingredient in producing a large extrinsic thermal Hall effect.

\begin{figure}[t]
    \centering
    \includegraphics[width=0.32\textwidth]{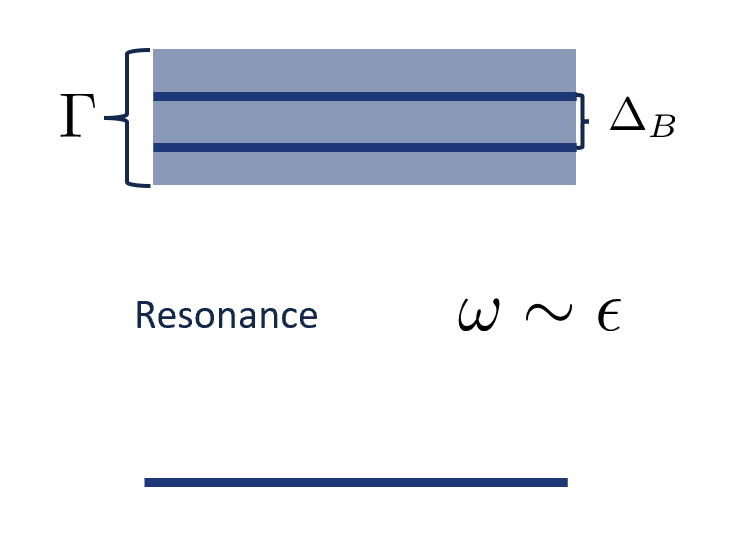}
    \caption{The three-level systems in resonance with a typical phonon energy $T$. Both intermediate states are in resonance if $|\Delta_B|<\Gamma< T$, where $|\Delta_B|$ is the level spacing induced by the magnetic field.} 
    \label{fig:three_level}
\end{figure}

In this paper, we explore the contribution to the thermal Hall effect from the resonant scattering of phonons 
from certain types of dynamical defects.  While a two-level system can lead to strong, resonate non-skew scattering~\cite{Anderson:1972,Phillips:1972}, it does not lead to  significant skew scattering. However, in the presence of a magnetic field, a three-level system with two nearly degenerate excited states~\footnote{As we will see, such three level systems require some form of effective local symmetry to produce the requisite degeneracy. Thus, this mechanism is expected to be pertinent only in a subset of insulators -- i.e. it must be less universal than those that arise from two-level systems.} leads to resonate scattering with a significant skew component. Specifically, the interference between virtual processes through the nearly degenerate intermediate states results in an enhancement of the skew scattering; in order for both processes to be simultaneously resonant, the  splitting of the two levels must be smaller than the widths of both intermediate states, as illustrated in Fig.~\ref{fig:three_level}. This is the key dynamical requirement for resonant skew scattering. However, a strong skew scattering is, by itself, not always sufficient to give rise to a significant phonon thermal Hall effect,  as has been noted in an important study~\cite{Mori:2014} of a different mechanism involving dynamical disorder (see below). There is also a key kinematic aspect. It turns out to be  important to take into consideration the breaking of rotation and inversion symmetries by \emph{some} individual scatterers -- even though the symmetries may be restored upon averaging.  

We explicitly evaluate  
$\kappa_H$ in a caricature model with physically reasonable parameters, and show it  
can account for 
a similar magnitude and $T$ dependence as observed in $\text{La}_2\text{CuO}_{\text{4}}$~\cite{Grissonnanche:2019}. Our caricature model involves both the usual non-skew scatterers,  about which we do not make detailed assumptions, as well as a certain kind of dynamical ionic defects as skew scatterers. The assumed dynamical defect involves a three-level structure consisting of a nearly degenerate doublet and a singlet as shown in Fig.~\ref{fig:three_level}, with an energy separation $\epsilon$. The doublet has a decay width $\Gamma$ and can be split by the magnetic field with a level spacing $\Delta_B$. We assume the energy separation $\epsilon$ is random with a distribution that is uniform over a range at least comparable to the relevant temperatures, $T \sim 10K$, which insures that they are resonant with the heat-carrying phonons.  On the other hand, the significance of $\Gamma$ and $\Delta_B$ enter through a \emph{resonance condition}, $|\Delta_B|<\Gamma$, which is needed for the aforementioned resonating interference to take place. As we will see below, the decay width $\Gamma$ of the dynamical defect under consideration, due to the emission of phonons, is proportional to $\epsilon^3$ times a constant that we will estimate explicitly.  The magnetic splitting $\Delta_B$ can be written as $\Delta_B = t e^{\star}Bb^2/\hbar c$,  where $b^2$ is the area enclosed in the ionic motion, $e^{\star}$ is an effective charge, and $t$ is a characteristic tunnelling energy. Since it is expected that $b$ cannot be more than a small fraction of an angstrom and $t$ must be no larger than the Debye energy, we will see that the resonance condition $|\Delta_B| < \Gamma$ is satisfied for typical values of the resonant energy $\epsilon$. In addition to their role in giving rise to the mean free path $\ell$, another principle role of the non-skew scatterers is that they randomly break rotation and inversion symmetry at the 
local level. We also discuss a few alternative sources for breaking the rotation and inversion symmetry of the skew scattering, including a scenario involving proximate pairs of scatterers discussed in Ref.~\cite{Mori:2014}.

\section{Boltzmann Transport}
In this work, we assume we are at sufficiently low temperatures that acoustic phonons are the major heat carriers, and the dominant scattering processes are elastic. By symmetry, $\kappa_H \equiv 0$ at zero magnetic field in many materials. The same symmetry considerations imply that the $\kappa_H$ grows linearly in an applied magnetic field $B$, where the extrinsic part of $\kappa_H$ comes from the magnetic field induced skew scattering. Skew scattering refers to components of the scattering rate that are time-reversal odd, i.e. that are antisymmetric under $\alpha\bk\leftrightarrow \alpha'(- \bk')$, where $\bk$ and $\bk'$ are the momenta of the incoming and outgoing phonons, and $\alpha$ and $\alpha'$ their polarization indexes. Since the phonon mean-free-path is much larger than the sound wavelength at the temperatures of interest, the semi-classical Boltzmann equation is suitable for studying transport. 

Our first step is to calculate the steady state phonon distribution in the presence of a small thermal gradient, $f_{l} = \bar{f}(\omega_l) + \delta f$, where $l\equiv \alpha\bk$ includes both phonon polarization and momentum, and $\bar{f}(\omega_l)$ is the unperturbed Bose-Einstein distribution as a function of phonon energy $\omega_{\bk}$. For small deviations from equilibrium, we can keep terms linear in $\delta f$ and in $ \nabla  T$, resulting in the linearized Boltzmann equation:
\begin{equation}
\begin{split}
    \frac{\partial \bar{f}(\omega_l)}{\partial T}\mathbf{v}_l\cdot\nabla  T = \sum_{l'} C_{ll'} \delta f_{l'}
\end{split}
\label{eq:Boltzmann}
\end{equation}
where the collision kernel is defined as
\begin{equation}
\begin{split}
C_{ll'}\equiv& \left[ (1+\bar{f}_{l}) W_{ll'} - W_{l'l} \bar{f}_l\right] \\
& - \delta_{ll'} \sum_{l''}\left[ (1+\bar{f}_{l''}) W_{l''l} - W_{ll''} \bar{f}_{l''}\right]
\label{W_to_C}
\end{split}
\end{equation}
with $W_{ll'}$ the scattering rate from the phonon state $l'$ to $l$. Since the scattering processes are elastic, $W_{ll'}$ and thus $C_{ll'}$ always contains a factor of $\delta(\omega_l-\omega_{l'})$.

As in Ref.~\cite{Chen:2020}, the scattering rate $W_{ll'}$ can be separated into a time-reversal even part, the non-skew scattering rate $W^e_{ll'}$, plus a time-reversal odd part, the skew scattering rate $W^o_{ll'}$ which is small in proportional to $B$, and  similarly, $C_{ll'}=C^e_{ll'}+C^o_{ll'}$. The variation of the phonon distribution $\delta f_l$ can also be separated into a time-reversal even part $\delta f^e_l$ of smallness $\nabla  T$, and a time-reversal odd part $\delta f^o_l$ of smallness $B\nabla T$. Treating $C_{ll'}$ as a matrix with indices $l$ and $l'$, at zeroth and first order in $B$ respectively, the linearized Boltzmann equation is solved by
\begin{equation}
    \begin{split}
        \delta f^e_l &= \sum_{l'} (C^e)^{-1}_{ll'}\frac{\partial\bar{f}(\omega_{l'})}{\partial T}\mathbf{v}_{l'} \cdot\nabla T, \\
        \delta f^o_l &= -\sum_{l',l''} (C^e)^{-1}_{ll'} C^o_{l'l''} \delta f^e_{l''} \\
        &=-\sum_{l'} \left[(C^e)^{-1} C^o (C^e)^{-1}\right]_{ll'} \frac{\partial\bar{f}(\omega_{l'})}{\partial T}\mathbf{v}_{l'}\cdot\nabla T.
    \end{split}
\end{equation}
The thermal current is given by $\mathbf{j}_Q = (1/V) \sum_l \mathbf{v}_l\omega_l \delta f_l$, with the $\delta f^e$ piece contributing the longitudinal current and the $\delta f^o$ piece contributing the Hall current. Matching with $\mathbf{j}^i_Q = -\kappa^{ij} \nabla_j T$, the longitudinal and transverse (Hall) conductivity tensors are
\begin{equation}
    \begin{split}
        \kappa_L^{ij} &= -\frac{1}{V}\sum_{l,l'} \mathbf{v}^i_l (C^e)^{-1}_{ll'} \mathbf{v}^j_{l'} \ \omega_l \frac{\partial\bar{f}(\omega_l)}{\partial T}, \\
        \kappa_H^{ij} &= \frac{1}{V}\sum_{l,l'} \mathbf{v}^i_l \left[(C^e)^{-1} C^o (C^e)^{-1}\right]_{ll'} \mathbf{v}^j_{l'} \ \omega_l \frac{\partial\bar{f}(\omega_l)}{\partial T}.
    \end{split}
\label{kappaH_expression}
\end{equation}
This is the more general version of Eq.~\eqref{kappaL_kappaH}, with $-(C^e)^{-1}_{ll'} \sim \tau$ understood in the relaxation time approximation. Henceforth, to simplify the discussion, we will assume that 
after averaging over disorder, the system has sufficient symmetry that $\kappa_L^{ij}=\kappa_L \delta^{ij}$ and $\kappa_H^{ij}=\kappa_H \epsilon^{ijk} \hat{B}^k$,  and we will define the direction of the magnetic field to be $\hat{z}$. In this case, the structure of the angular integral involved above implies that one may think of the thermal Hall effect as coming from the component in $\left[(C^e)^{-1} C^o (C^e)^{-1}\right]$ which effectively behaves as $\mathbf{B}\cdot (\bk\times\bk')$, or, in terms of the inclination and azimuth angle $(\theta_\bk, \phi_{\bk})$ of the direction of momentum with respect to the $\mathrm{xy}$-plane, the $j=|j_z|=1$ component proportional to $\sin\theta_\bk \sin\theta_{\bk'} \sin(\phi_{\bk}-\phi_{\bk'})$. Naturally, there can be higher orbital angular momentum terms in the skew scattering $C^o$. This is a point that will turn out to be important in the below. 

Before we move on, we would like to make a remark on anti-reciprocality versus skewness. A component of the scattering rate is reciprocal or anti-reciprocal if it is even or odd under $l \leftrightarrow l'$. Comparing this to the definition of skewness, we see that skew scattering is anti-reciprocal if and only if the process is invariant under inversion $\bk\leftrightarrow -\bk$ (and the same for $\bk'$ simultaneously).

\section{Dynamics of Resonant Skew Scattering}

\begin{figure}[t]
    \centering
    \includegraphics[width=0.44\textwidth]{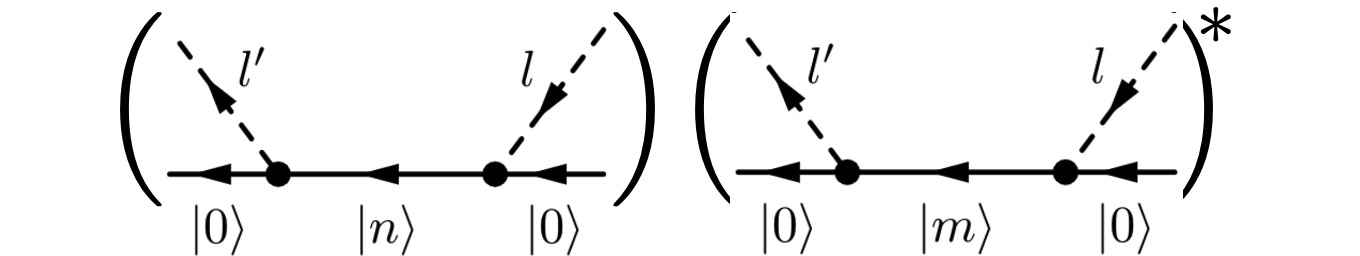}
    \caption{One of the interference terms between two processes that contribute to resonant skew scattering rate.}
    \label{fig:diagram}
\end{figure}

Now we discuss how \emph{dynamical} defects can give rise to large skew scattering rates $W_{l'l}^o$ due to resonance effects. In this situation, the scattering of phonons is mediated by virtual transitions in which the defect absorbs and emits a phonon. Suppose the initial and final state of the defect are its ground state $|0\rangle$, whose energy is set to $0$. 
(The resonance analysis below also applies when the initial and final state is, instead, a thermally occupied excited state.) 
As justified in Appendices \ref{app_estimation} and \ref{app_formal}, the scattering rate, $W_{l'l}= 2\pi\delta(\omega_l-\omega_{l'}) |T_{l'l}|^2$, is determined by the square of a scattering amplitude given by  the T-matrix which sums over all possible intermediate states of the defect and is given by
\begin{equation}
\begin{split}
    T_{l'l}=& \ \sum_{n}U_{n,l'}^{*}\frac{1}{\omega-E_n+i\Gamma_n/2}U_{n,l} \\ & \ + \sum_{n} U_{n,l}\frac{1}{-\omega-E_n+i\Gamma_n/2} U_{n,l'}^{*}.
\end{split}
\end{equation}
Here $\omega=\omega_l=\omega_{l'}$ is the incoming/outgoing phonon energy, $E_n$ and $\Gamma_n$ are the energy and line width of the intermediate defect state $|n\rangle$, and $U_{n,l}$ is an interaction matrix element between the defect and phonon, $U_{n,l}=\langle n |H_{int}|0,l\rangle$. The first term is the process of absorbing the phonon $l$ first and emitting the phonon $l'$ later, while the second term is the process with the opposite order. Note only the first term can be resonant, so in the below we will ignore the second term. For the kind of resonance mechanism that we will introduce in this paper, the resonant skew component is symmetric under inversion, so that it is also anti-reciprocal, i.e. odd under $l\leftrightarrow l'$, as remarked before. In this case, the skew, anti-reciprocal component in $|T_{l'l}|^2$ comes from antisymmetrizing the expression between $l$ and $l'$, which only receives contributions from the interference between a pair of \emph{distinct} intermediate states, as shown diagrammatically by Fig.~\ref{fig:diagram}:
\begin{equation}
\begin{split}
    |T_{l'l}|^2_o = & \sum_{n\neq m} \frac{(E_{m}-E_{n})(\Gamma_m+\Gamma_n)/2+(\cdots)}{\left[(\omega-E_n)^2+\Gamma_n^2/4\right]\left[(\omega-E_{m})^2+\Gamma_{m}^2/4\right]}\\
    &\cdot \frac{i}{2}\left(U_{n,l'}^{*}U_{n,l}U_{m,l}^{*}U_{m,l'}-U_{n,l}^{*}U_{n,l'}U_{m,l'}^{*}U_{m,l}\right),
\end{split}
\label{eq:T-square}
\end{equation}
where $(\cdots)= (\Gamma_m-\Gamma_n)(2\omega- E_{m}- E_{n})/2$. The skew scattering rate is highly modulated by the energy denominator. In particular, if a pair of states $|n\rangle$ and $|m\rangle$ have nearly degenerate energies, $0<|E_{m}- E_{n}|\lesssim \Gamma_n$, then when a phonon comes in with energy $\omega$ such that $|\omega- E_n|\lesssim\Gamma_n$, both factors in the denominator are simultaneously small (of order $\Gamma^2$) -- i.e. at resonance. In this case, the $(\cdots)$ term in Eq.~\eqref{eq:T-square} can be neglected compared to  the first term.

For a system that is time-reversal symmetric to begin with, an applied magnetic field can produce skew scattering both through the field dependence of the energy levels and of the transition matrix elements. If $|n\rangle$ and $|m\rangle$ are a time-reversal pair of degenerate states whose degeneracy is only lifted by the magnetic field, $|E_m- E_n|=|\Delta_B|\propto B$, then in Eq.~\eqref{eq:T-square}, to leading order in $B$, we can neglect the $B$ appearance in the transition matrix elements, and the resonance condition is $|\Delta_B|<\Gamma$. If, instead, the zero field degeneracy between the two excited states is already weakly lifted in a manner that respects time-reversal symmetry, then the energy splitting $E_m-E_n$ is unaffected by $B$ to linear order, but the $B$ dependence will reappear in the transition matrix elements in such a way that the dominant contribution is comparable to the previous degenerate case, as long as both  $|E_m- E_n|<\Gamma$ and $|\Delta_B|<\Gamma$; in particular, it does not require $|E_m- E_n|<|\Delta_B|$. (See Appendix~\ref{app_nearly_deg} for details.) In the below, therefore, it suffices to consider the degenerate case.

A remarkable feature of our mechanism is that, as long as the $|\Delta_B|<\Gamma$ resonance condition is satisfied, the skew scattering rate is to leading order independent of the strength of the defect-phonon coupling. This can be seen from the energy denominator in Eq.~\eqref{eq:T-square} when the two energies are nearly degenerate at $E$ :
\begin{equation}
    W_{l'l}^o \sim \frac{\Delta_B\Gamma g^4}{[(\omega-E)^2+\Gamma^2/2]^2}\sim \frac{\Delta_B g^4}{\Gamma^2}\delta(\omega-E),
\end{equation}
where $g$ is the defect-phonon coupling, the factor of $g^4$ in the numerator reflects the fact that the $T$ matrix 
is second order, and $\delta(\omega- E)$ is to be interpreted as a normalized peaked function of width $\sim\Gamma$ and height $\sim 1/\Gamma$. Compared to the familiar resonant non-skew scattering rate \cite{Anderson:1972,Phillips:1972}, this resonant skew scattering rate is smaller by a factor of $|\Delta_B|/\Gamma$. Assuming $\Gamma$ is dominated by phonon-assisted decay, we have $\Gamma \sim g^2$ (see Appendix~\ref{app_formal} for theoretical justification), and hence the frequency integrated skew scattering rate is independent of the magnitude of the defect-phonon coupling to leading order, so long as $|\Delta_B|<\Gamma$ remains valid. 

At this point we wish to remark on the crucial difference between the present mechanism and that in the pioneering work Ref.~\cite{Mori:2014}, despite some apparent resemblances. Our resonance enhancement mechanism crucially relies on the condition $|E_m- E_n|\sim |\Delta_B| < \Gamma$, so that both intermediate states are simultaneously at resonance while interfering. In contrast, in Ref.~\cite{Mori:2014}, even in the absence of magnetic field $|E_m- E_n|\gg \Gamma$; moreover the assumed splitting due to magnetic field in their model satisfies the opposite inequality, $|\Delta_B|\gg \Gamma$. As a consequence, at most one of the two interfering intermediate states can be at resonance at a time. As we will see below, this crucial difference in the energetics is closely related to the difference of the underlying pictures of the origin of the dynamic impurities -- in their case associated with interstitial magnetic ions and in our case associated with tunnelling of ionic structural defects.

\section{Kinematic considerations}

The resonant skew scattering mechanism we have discussed solves the ``big'' dynamical problem - providing a clear and simple mechanism by which an applied magnetic field can significantly affect the phonon dynamics. We now turn to some aspects of the consequent kinematics of the phonons, which have important implications for the thermal Hall effect. 

The simplest example of the kind of dynamical defect introduced above is a three-level defect with a time-reversal invariant ground state $|0\rangle$ and a pair of excited states $|n\rangle$ and $|\bar{n}\rangle$ that are related by time-reversal transformation, with splitting $\Delta_B\propto B$. This is illustrated in Fig.~\ref{fig:three_level}; no further detail of the defect is needed. We are now interested in the angular distribution of the phonons involved in the scattering process. The key observation, as shown diagrammatically in Fig.~\ref{fig:diagram}, is that the dependence of the resonance scattering rate on the incoming wave direction $\hat{\bk}$ appears only through the factor $U_{n,l}U_{\bar{n},l}^{*}$ and its complex conjugate. For the matrix element $U_{n,l}$, we decompose it into the symmetric and antisymmetric parts under inversion $\bk\rightarrow -\bk$:
\begin{equation}
\begin{split}
    U_{n,l}&= U_{n, l}^S + U_{n, l}^A.
\end{split}
\end{equation}
In terms of spherical harmonics, $U^S$ and $U^A$ respectively contains all the components with even and odd orbital angular momenta $j$. Time-reversal takes $|n\rangle$ to $|\bar{n}\rangle$ and $\bk$ to $-\bk$, and is anti-unitary, therefore
\begin{equation}
\begin{split}
    U_{\bar{n},l}^\ast&= U_{n, l}^S - U_{n, l}^A,
\end{split}
\end{equation}
and consequently,
\begin{equation}
\begin{split}
    U_{n,l} U_{\bar{n},l}^\ast&= (U_{n, l}^S)^2 -(U_{n, l}^A)^2,
    \label{res_skew_k_dependence}
\end{split}
\end{equation}
which is, remarkably, invariant under $\bk\rightarrow -\bk$~\footnote{We emphasize again that this conclusion is restricted to the intermediate states being only the simultaneously resonant time-reversal pair of states. It does not apply to the non-resonant -- and hence much smaller -- contributions to the skew scattering rate.}. The same result applies to the outgoing phonon direction $\hat{\bk}'$. This means that the resonant -- and hence dominant -- contribution to the skew scattering rate $W^o_{l'l}$ is not only invariant under simultaneously reversing $\bk$ and $\bk'$, but more strictly, it must be invariant under reversing either one of $\bk$ 
or $\bk'$. In terms of spherical harmonics, only even $j$ are included in the resonance contribution to $W^o$. A simple example of such angular dependence that can appear in the skew scattering rate $W^o$ is
\begin{align}
    & iY_2^{+2}(\theta_{\bk'}, \phi_{\bk'}) Y_2^{-2}(\theta_\bk, \phi_\bk) - iY_2^{-2}(\theta_{\bk'}, \phi_{\bk'}) Y_2^{+2}(\theta_\bk, \phi_\bk) \nonumber \\ \propto & \: \sin^2\theta_\bk \, \sin^2\theta_{\bk'}\, \sin[2(\phi_\bk-\phi_{\bk'})]
    \label{Wo_form}
\end{align}
where $Y_j^{j_z}$ are the spherical harmonics. This particular form of angular dependence is what we will encounter in our caricature model below. (This same factor  also appeared, for similar kinematic reasons, in the already mentioned model in Ref.~\cite{Mori:2014}, despite its distinct dynamical structure.)

This general analysis has important implications in the context of the thermal Hall effect. From Eq.~\eqref{kappaH_expression} we concluded that the thermal Hall effect arises from the $j=|j_z|=1$ skew component of $\left[ (C^e)^{-1} C^o (C^e)^{-1}\right]_{ll'}$. While this component is invariant under rotation (fixing the magnetic field along $\hat{z}$) and inversion, it is not invariant under reversing either one of $\bk$ or $\bk'$ individually. As we discussed above, however, this is a property of the resonant contribution to $C^o$.

Consequently, in order for resonant skew scattering to generate a significant thermal Hall effect, we cannot merely use the relaxation time approximation which replaces the matrices $(C^e)^{-1}$ by a constant $\tau$.  Instead, we must invoke suitable inversion and rotation breaking contributions (since either inversion or rotation can reverse a momentum) to the non-skew processes encoded in $(C^e)^{-1}$. Physically, such contributions are expected to exist at the individual defect level. 
An alternative way to produce a significant thermal Hall effect is to modify the assumption to begin with, so that the two interfering intermediate states $
|n\rangle$ and $|\bar{n}\rangle$ are not time-reversal pairs of each other; but then there must be some other reason why the states are nearly degenerate. In the below we elaborate on both possibilities.

We first consider the scenario of invoking the inversion and rotational symmetry breaking contributions in $(C^e)^{-1}$. For concreteness we consider the explicit (simplest) angular dependence of the resonant skew scattering rate 
consistent with the previous arguments, i.e. we take $W^o_{l'l}$ of the form given in Eq.~\eqref{Wo_form}. (More general cases work in a similar manner.) Since $W^o$ involves $j=|j_z|=2$ only, while only $j=|j_z|=1$ contribute to $\kappa_H$, we need certain local non-skew scattering defects that break rotation and inversion symmetry in such a way that $W^e$ mixes $j=|j_z|=2$ and $j=|j_z|=1$. We denote such terms as $W^e_{break}$. Since such terms do not have rotational symmetry, they will depend on the local orientations of the associated defects (see below for more discussions). The details of the terms and the calculations are presented in Appendix \ref{app_invert}. Ignoring phonon polarization (since it will only make order $1$ differences), we find
\begin{equation}
    \begin{split}
     \kappa_L &= \int d\omega \, \mathcal{D} \, \frac{v^2}{3} \omega \, \tau \frac{\partial\bar{f}}{\partial T} \ , \\
     \kappa_H &= \int d\omega \, \mathcal{D} \, \frac{v^2}{3} \omega \, \left(\gamma w^o\tau\right) \tau \left( 1+2\bar{f}\right)^3 \frac{\partial\bar{f}}{\partial T} \ ,
    \label{eq:kappa_H}
    \end{split}
\end{equation}
where $\mathcal{D}=4\pi\omega^2/(2\pi v)^3$ is the density of states, $\tau(\omega)$ is an effective relaxation time that contributes to the longitudinal transport, $w^o(\omega)\equiv\tau_o^{-1}$ is the coefficient of $W^o$ normalized to have dimension [1/time], and $\gamma$ is a dimensionless ratio that captures the typical size of the rotation and inversion breaking effects:
\begin{align}
    \gamma \sim \frac{(W^e_{break})^2}{(W^e)^2} \ .
\end{align}
The dependence on the local orientation of the rotation and inversion breaking defects turns out to drop out. Compared to $\kappa_L$, the integrand for $\kappa_H$ involves an extra dimensionless factor $\gamma w^o \tau (1+2\bar{f})^3$, which is similar to the expression in Eq.~\eqref{kappaL_kappaH}, but with the new ingredient $\gamma$. Since the total scattering probability $W^e_{l'l}$ must be $\geq 0$ for any $l'l$, by construction $|\gamma|$ cannot generally be larger than order $1$. (On the other hand, the overall sign of $\gamma$ is undetermined, since different terms in $W^e_{break}$ may contribute to $\gamma$ with different signs, see Appendix \ref{app_invert}.) 

Let us discuss the physical nature of this factor $\gamma$ in more detail. In typical materials the orientations of the rotation and inversion breaking defects are expected to be locally random. In order for these orientations to have well-defined local values in the Boltzmann equation (through the expressions of $W^e_{break}$, see Appendix \ref{app_invert}), however, the local randomness of these orientations must be correlated within a scale that is at least as large as the phonons' typical wavelength, because the notion of ``position'' in the Boltzmann equation is, indeed, only well defined down to the scale of the particle's quantum wavelength. Since there are two factors of $(C^e)^{-1}$ in $\kappa_H$, we expect the local correlation to give rise to a non-vanishing result after averaging over the locally random orientations, which restores rotational symmetry -- this is reflected by the fact that the dependence on the local orientations drops out in the contribution to $\kappa_H$. Therefore $\gamma$ should really be thought of as some sort of fractional co-variance, taken within one phonon wavelength, of such rotational symmetry breaking scattering effects. If the density of the relevant scatterers is $\lesssim 1$ within one phonon wavelength, then the co-variance $\gamma$ becomes a defect's self variance, which is unlikely to be small.  Alternatively, if the scatterers are extended defects, again $\gamma$ need not be small.

The non-skew scattering is not the only possible source of the breaking of the rotational and inversion symmetries required by the kinematics. An alternative possibility was suggested in Ref.~\cite{Mori:2014}. Suppose the skew scatterers are still three-level defects, but consider the case in which their energy levels are given, i.e. not randomly distributed. Then there can arise a new kind of resonant interference: the $|n\rangle$ state on one defect can interfere with the $|\bar{n}\rangle$ state on a nearby defect. However, for this mechanism to survive averaging over a random spatial distribution, the spatial distribution of the defects must be correlated, i.e. given the position of one, the probability of finding another, $p(\delta {\bf r})$, depends on its displacement $\delta{\bf r}$ from the first.  Specifically, what is required is that $\int_{\delta\br} p(\delta\br) e^{i(\bk-\bk')\cdot\delta\br} \neq 0$ for $\bk\neq\bk'$. In this scenario, the interfering \emph{pair} of defects again breaks rotation symmetry locally (although the symmetry is restored upon averaging), and the scale of this integral plays the role of $\gamma$ in our preferred scenario. While this mechanism may be relevant to the system considered in Ref.~\cite{Mori:2014}, it does not seem likely to be relevant to the systems considered here.  In particular,  the assumption of \emph{identical} defects implies that the thermal Hall current must consist almost entirely of the phonons whose frequencies are within a narrow window of width $\Gamma$ around $E_n$, which is a small fraction of all phonons.  As far as we can determine, for reasonable parameters this small fraction of phonons cannot carry a large enough thermal Hall current to give an effect comparable to the observed ones in the materials of interest.

Note that there are also systems with different symmetries in which these kinematic considerations do not arise. For instance if a lattice has $C_3$ symmetry in the $\mathbf{xy}$-plane, then we would expect some defects, whose preferred orientation aligns with the lattice modulo $2\pi/3$, have non-vanishing matrix elements with $j_z=\pm3$, and hence can scatter phonons between $j_z=+2$ and $j_z=-1$ and between $j_z=-2$ and $j_z=+1$. In such cases, $\gamma$ does not need to be understood in the sense of a local co-variance, as the orientation is not random.

\section{An illustrative model}
To facilitate a more quantitative analysis, we construct an explicit model of the aforementioned three-level defects. We consider a dipolar defect consisting of a localized positive charge and a dynamical negative charge that can occupy any of the four sites forming a square in the $xy$-plane (assuming $\bB= B\hat{z}$). The dynamics of the dipolar defect is governed by the effective Hamiltonian
\begin{equation}
   H_{dip}= -\sum_{i>j}t_{ij} \hat H_{ij},\quad  \hat H_{ij} \equiv \left[c_{\mathbf{R}_i}^{\dagger}c_{\mathbf{R}_j}e^{ie^\star A_{ij} }+ h.c.\right],
   \label{eq:Ham_dip}
\end{equation}
where the site index $i=1,..,4$ and $\mathbf{R}_i$ labels the possible orientations of the dipole, and $A_{ij}=- A_{ji}$ is the integral of the vector potential along the bond connecting site $j$ to site $i$.
( In common with Refs.~\cite{Anderson:1972,Phillips:1972}, we will not attempt to pin down the microscopic nature of this defect. But for visualization purposes one might think of the four states as representing different tilt angles of an oxygen pyramid containing a positive charged metal ion~\footnote{There are a variety of theoretical and empirical reasons to believe that there are generically some form of ``local dipolar modes'' in underdoped cuprates -- for an early review see Ref.~\cite{physicaC}.}). Consistent with the symmetry of the square, we take $t_{ij}$ to be $t$ and $t'$ between nearest and next-nearest neighbor sites, respectively. Energy eigenstates are labeled by the orbital ``angular momentum'', $\ell \ \text{mod}\: 4$, around the square. Keeping up to linear order in $B$, the state with $\ell=0$ has energy $E_0= - 2t- t'$, those with $\ell=\pm 1$ have $E_\pm = t'\mp \Delta_B/2$ where $\Delta_B = t e^{\star}Bb^2$ ($b$ is the distance between nearest-neighbor sites), and that with $\ell=2 \cong -2$ has $E_2 = 2t- t'$. We consider the case in which $t>0$, $t'< 0$ and such that, in the absence of a magnetic field, $\epsilon \equiv E_\pm - E_0=2(t-|t'|)$ is small compared to $E_2- E_0=4t$. In this case, the highest state can be neglected, and the system can be well approximated as a three-level system with one singlet $|0\rangle$ and one doublet $|\pm \rangle$, with the latter split only by the magnetic field. Note that $\epsilon$ can take either sign, i.e. $|0\rangle$ might have lower or higher energy than $|\pm\rangle$. We will assume that our three-level systems have a random distribution in $\epsilon$, such that the density of the three-level systems per unit energy per unit volume, $n(\epsilon)$, is broad near $\epsilon=0$, similar to Refs.~\cite{Anderson:1972,Phillips:1972}. (We can assume the broad distribution extends over some tens of Kelvins.)  They thus contribute a linear in $T$ term to the specific heat at low temperatures. 

Now we turn to the acoustic phonons. They are described by the acoustic displacement $\bu$ and the conjugate momentum $\bPi$; at low temperatures we may assume the phonons are non-interacting among themselves. Since acoustic phonons are Goldstone modes, they may only have derivative couplings to the defect Hamiltonian -- either through spatial gradients (i.e. strain) or time derivatives (i.e. $\dot{\bu} =\bPi/\rho $).  To linear order in gradients, and consistent with time-reversal symmetry at $B=0$, these couplings are of the form
\begin{align}
    H_{int}=&  
    \frac{1}{4}[\partial_a u_b+\partial_b u_a]\ \tilde g_{ij}^{ab}\ \hat H_{ij}
     + \frac{1}{2}[\Pi_a /\rho] \ g_{ij}^{a} \ \hat J_{ij},
     \label{eq:Ham_int}
\end{align}
\begin{align}
    \hat J_{ij}  \equiv i[c_{\mathbf{R}_i}^\dagger c_{\mathbf{R}_j}-h.c.],
    \label{eq:current}
\end{align}
where $a, b= x, y, z$, and summation over repeated indices is understood. The assumed $C_4$ symmetry of the system imposes additional constraints on the form of the coupling constants, $g$ and $\tilde g$. It turns out the strain coupling $\tilde g$ is unimportant at resonance so we may set $\tilde g=0$~\footnote{The $\tilde{g}$ coupling only changes the $C_4$ ``angular momentum'' $\ell$ by $2 \  \text{mod} \: 4$. The transition between $\ell=0$ and $\ell=2$ is unimportant because the energy change is too large for thermal phonons to resonate with. The transition between $\ell=\pm 1$ is unimportant because the energy change is too small, so that the resonating phonons -- those of very large wavelengths -- have tiny phase space and, moreover, small strain.}.
On the other hand, for any pair of nearest neighbor sites $i$ and $j$, 
\begin{align}
   g^a_{ij} = g\ (R^a_i - R^a_j)/b
   \label{eq:coupling}
\end{align}
i.e. it represents the symmetry allowed coupling between the electric current, $\hat J$, and the phonon velocity. For simplicity, we take $g^a_{ij}=0$ for further neighbor sites -- this has no qualitative effect on our results. In order to ensure the resonance condition $|\Delta_B|<\Gamma$ is met, we need an estimate of $g$ based on its microscopic origin, and it turns out $g\sim \hbar/b$ (see Appendix~\ref{app_g_origin}).  

Details of calculations with this model are presented in Appendix~\ref{app_model_3-level}. Here we reiterate two generic features of our mechanism. The first is that the thermal Hall effect thus generated is to leading order independent of $g$. The second is, even if the defect is subjected to some small random electric potential, so that the four sites have slightly different potential energies of order $\Delta_E$, the thermal Hall effect is largely unaffected as long as $|\Delta_E|<\Gamma$, regardless of the relative size between $|\Delta_B|$ and $|\Delta_E|$.

Now we demonstrate that for reasonable model parameters (including plausible assumptions about the requisite inversion symmetry breaking aspects of the non-skew scattering needed to satisfy kinematic constraints), this model can produce a magnitude of the thermal Hall effect that is comparable to those observed in recent experiments. (Note we make \emph{no} claim that our present model is a valid microscopic model of any particular material.) We can estimate the thermal Hall conductivity from Eq.~\eqref{eq:kappa_H}. Generally, $\tau$ and $\gamma$  depend on the phonon energy. Nonetheless, since we are only trying to get a intuition for magnitudes, we take them to be constants representing some appropriate average.  The thermal Hall conductivity can then be estimated as  
\begin{equation}
    \kappa_H\sim \gamma n(0)\Delta_B T (v\tau)^2.
\end{equation}
For the assumed three-level defect, we take dimensionally reasonable values: $t=100 \mathrm{K}$, $e^\star=- 2|e|$, $b=0.1\Ang$ and $g=\hbar/b$, i.e. the main hopping body is assumed to be an oxygen ion, with a tunneling energy $t$ somewhat smaller than a typical optical phonon energy, and the range $b$ is the largest  distance it might reasonably be expected to tunnel. This gives $|\Delta_B|\approx0.0005\mathrm{K}$ at $15\mathrm{T}$, meaning that the resonant condition $|\Delta_B|<\Gamma$ is satisfied for defects with level spacing $|\epsilon|\gtrsim 2.5\mathrm{K}$, where the estimation of $\Gamma$ is at $T=10\mathrm{K}$ and we also consider a typical phonon velocity of $v=300\mathrm{K \AA}$ ($\approx 3900 \mathrm{m/s}$), mass density of $\rho=0.087 \mathrm{K^{-1}\AA^{-5}}$ ($\approx 7000 \mathrm{kg}/\mathrm{m}^3$). Thus, if we take $v\tau\sim 3\mathrm{\mu m}$, around $T=10\mathrm{K}$ and $B=15\mathrm{T}$, we need a $|\gamma| n(0)\sim 1.3\times 10^{-7}/(\mathrm{K}\Ang^3)$ (see below for a discussion) to produce a thermal Hall effect of the same order of magnitude as that observed in La$_2$CuO$_4$, $\kappa_H\sim 10\mathrm{mW/K\cdot m}$. For these parameters, the Hall ratio can be estimated as 
\begin{equation}
    \frac{\kappa_H}{\kappa_L}\sim\frac{\gamma n(0) \Delta_B v^3 \tau}{T^2}\sim 2\times10^{-3},
\end{equation}
which is compatible with the recent experiments \cite{Chen:2021}.

Since the assumed resonant scatters would contribute an amount $C_{res}\approx 1.9 n(0)T$ to the low temperature  specific heat, an important consistency check is to make certain that this contribution does not exceed measured values. In the specific heat measurement of undoped La$_2$CuO$_4$, the fitted linear in $T$ part of the specific heat has a slope likely less than $1\mathrm{mJ}/(\mathrm{K}^2\cdot\mathrm{mol})$~\cite{Komiya:2009,Girod:2021}. This gives an upper bound of $n(0)$ of around $7\times 10^{-7}/(\mathrm{K}\Ang^3)$. A  value of $n(0)$ below this estimated bound can still produce a large enough $\kappa_H$, but only if we assume $\gamma$ is of order $1$ (recall $|\gamma|$ cannot be larger than order $1$ in typical scenarios, by construction). A possible way to achieve a $\gamma$ of order $1$ is that the non-skew scattering is dominated by scattering of phonons by extended structural defects so that the shape of the defects remains relevant in the low energy scattering. 

A second consistency check involves the contribution to the non-skew  scattering rate of resonant scattering by the three-level defects themselves.  This must be smaller than the assumed full scattering rate $1/\tau$, since our assumptions concerning the kinematics require  that $\tau$ reflects the presence of other scatterers. The scattering rate of a typical phonon of energy $T$ off the three-level defects can be estimated as: $1/\tau_{3-level} \sim n(0)g^2T/\rho\sim 0.003 \mathrm{K}$ for $n(0)=3\times 10^{-7}/(\mathrm{K}\cdot \Ang^3)$ and $T=10\mathrm{K}$. For comparison, the value of $1/\tau$ that gives a mean free path of $3\mathrm{\mu m}$, a value that gives a comparable $\kappa_L$ with the measured values, is $0.01\mathrm{K}$. 

\section{Final thoughts}
The fact that phonons can contribute to the thermal Hall effect may have implications for the interpretation of a variety of interesting recent observations in insulators. For instance a record-setting thermal Hall conductance has been reported~\cite{Chen:2021} in Cu$_3$TeO$_6$. 

One particularly striking feature of the thermal Hall effect in the cuprates~\cite{Grissonnanche:2019,Grissonnanche:2020,boulanger2020thermal} is that an identifiable phonon contribution persists when the system (e.g. $\text{La}_2\text{Cu}\text{O}_4$) is doped.  The phonon $\kappa_H$ decreases monotonically with increasing doping $p$ across the range of doping comprising the ``superconducting dome,'' and only vanishes (or becomes undetectably small) for $p> p^\star$ where $p^\star$ marks the end of the so-called pseudo-gap regime.   On the theoretical front, many new processes are allowed in a metallic system directly related to the electronic Hall effect, and possibly indirectly related to the scattering of phonons from electrons.  It is beyond the scope of the present paper to analyze all these possibilities.  However, one  effect that  arises in the context of the present model is that impurity-electron coupling opens a new channel for the decay of the excited states in the three-level systems.  Now, there is a totally new source of decay that is proportional to the electron density of states at the Fermi energy $\rho$ and the impurity-electron coupling constant squared, $g_{el}^2$.  Since the skew scattering rate is proportional to $\Gamma^{-2}$, this leads to a suppression of $\kappa_H$ in a metallic state proportional to $[ g_{el}^2\rho ]^{-2}$ which could lead to a rapid quenching of the effect as the pseudo-gap disappears.

{\em Note added:}  As we neared completion of this work, a paper by Flebus and MacDonald was posted addressing the same issues~\cite{macdonald}. In agreement with our analysis, they concluded that the large thermal Hall effect must be extrinsic, associated with skew scattering of phonons from defects.  In contrast to the present results, they considered charged non-resonant defects that move in the acoustic field independently of any nearby compensating charge. During the review process, we became aware of another paper by Guo, Joshi and Sachdev~\cite{Guo:2022}  in which another possible resonance effect in the thermal Hall transport is discussed.

\begin{acknowledgments}
\emph{Acknowledgments.--} We thank John Tranquada for discussions on structural defects in cuprates, Louis Taillefer and Mohit Randeria for useful comments. XQS was supported by the Gordon and Betty Moore Foundations EPiQS Initiative through Grant No.~GBMF8691. JYC was supported by the NSFC under Grants No.~12174213 and No.~12042505. SAK was supported, in part, by the Department of Energy, Office of Basic Energy Sciences, under Contract No. DEAC02-76SF00515 at Stanford. 
\end{acknowledgments}

\appendix
\begin{widetext}

\section{Estimation of Temperature Scaling for Resonance Scattering}
\label{app_estimation}

In this appendix we summarize how to estimate the temperature scaling for both non-resonant and resonant scattering. This will explain why we have chosen to explore the dynamical mechanism that we introduce in the main text, as opposed to other possibilities that involve magnetic field and resonance. We believe such kind of estimations is not only helpful for our present work, but also generally useful for future studies in this field. 

It is standard in the computation of resonance scattering to assume that:
\begin{enumerate}
\item The resonance width of the dynamical defect is dominated by the one phonon loop process.
\item The collision kernel is dominated by the one phonon in, one phonon out process.
\end{enumerate}
These assumptions can be justified in perturbation theory, as we will explain in Appendix \ref{app_formal}. For now let us follow the standard practice and summarize the estimation. 

The collision kernel, according to the second assumption above, is dominated by the $1$-to-$1$ scattering T-matrix square, given by
\begin{eqnarray}
\left|T^{fi}_{l'l} \right|^2 \ \ = \ \ \left| \ \ \ \ \sum_{|n\rangle} \ \ \left(
\parbox{40mm}{
\begin{center}
\begin{fmffile}{zzz-schmematic1}
\begin{fmfgraph*}(30, 10)
\fmfstraight 
\fmfleftn{l}{2}\fmfrightn{r}{2}
\fmfset{arrow_len}{2.8mm}
\fmfset{dash_len}{2.5mm}
\fmf{plain_arrow, tension=2, label=$|i\rangle$, label.side=left, width=1.8}{r1,m2}
\fmf{plain_arrow, tension=1, label=$|n\rangle$, label.side=left, width=1.8}{m2,m1}
\fmf{plain_arrow, tension=2, label=$|f\rangle$, label.side=left, width=1.8}{m1,l1}
\fmffreeze
\fmf{dashes_arrow, label=$l$, label.side=right, label.dist=4}{r2,m2}
\fmf{dashes_arrow, label=$l'$, label.side=right,label.dist=4}{m1,l2}
\fmfdot{m1}
\fmfdot{m2}
\end{fmfgraph*}
\end{fmffile}
\end{center}
}
+
\parbox{40mm}{
\begin{center}
\begin{fmffile}{zzz-schmematic2}
\begin{fmfgraph*}(30, 10)
\fmfstraight 
\fmfleftn{l}{2}\fmfrightn{r}{2}
\fmfset{arrow_len}{2.8mm}
\fmfset{dash_len}{2.5mm}
\fmf{plain_arrow, tension=2,label=$|i\rangle$,label.side=left, width=1.8}{r1,m2}
\fmf{plain_arrow, tension=1,label=$|n\rangle$,label.side=left, width=1.8}{m2,m1}
\fmf{plain_arrow, tension=2,label=$|f\rangle$,label.side=left, width=1.8}{m1,l1}
\fmffreeze
\fmf{dashes}{rm2,m1}
\fmf{dashes_arrow, label=$l$, label.side=right, label.dist=3}{r2,rm2}
\fmf{dashes,tension=2}{m2,lm2}
\fmf{phantom,tension=3}{lm2,llm2}
\fmf{dashes_arrow, label=$l'$, label.side=right, label.dist=3}{llm2,l2}
\fmfdot{m1}
\fmfdot{m2}
\fmffreeze
\end{fmfgraph*}
\end{fmffile}
\end{center}
}
\right) \ \ \right|^2
\\ \nonumber
\end{eqnarray}
where the dashed lines are the incoming and outgoing phonons $l=\mathbf{k}\alpha$ and $l'=\mathbf{k}'\alpha'$, and the solid line the dynamical defect in its initial, intermediate, and final states, with the intermediate state summed over. Therefore, the magnitude of a typical term in $|T^{fi}_{l'l}|^2$ is estimated as
\begin{align}
\sim U_{in} U_{out} U_{in'} U_{out'} G_n G_{n'} \ .
\end{align}
Here $G_n$, represented by the thickened line, is the interacting full Green's function of the defect's intermediate state, parametrized as
\begin{align}
G_n = \frac{1}{E - E_n + i\Gamma_n/2}
\end{align}
with $E=E_i+\omega_l$ or $E=E_i-\omega_{l'}$ in the diagrams above; since the incoming states are from a thermal ensemble, we typically consider the thermal activation energy $E-E_0 \sim T$, where $E_0$ is the ground state energy of the defect. The width $\Gamma_m$ is usually small and unimportant for most values of $E$, except when $E$ and $E_m$ are close, i.e. near resonance. Near resonance we need to estimate $\Gamma_m$, which, according to the first assumption above, is dominated by
\begin{eqnarray}
\Gamma_n/2 \ \ = \ \ \mathrm{Im}\left[ \ \ \ \ \sum_{|m\rangle} \ \ \left(
\parbox{40mm}{
\begin{center}
\begin{fmffile}{zzz-schmematic3}
\begin{fmfgraph*}(30, 8)
\fmfstraight 
\fmfleftn{l}{2}\fmfrightn{r}{2}
\fmfset{arrow_len}{2.8mm}
\fmfset{dash_len}{2.5mm}
\fmf{plain_arrow, tension=2,label=$|n\rangle$,label.side=left, width=1}{r1,m2}
\fmf{plain_arrow, tension=1,label=$|m\rangle$,label.side=left, width=1}{m2,m1}
\fmf{plain_arrow, tension=2,label=$|n\rangle$,label.side=left, width=1}{m1,l1}
\fmffreeze
\fmf{dashes_arrow,right=1}{m2,m1}
\fmfdot{m1}
\fmfdot{m2}
\fmffreeze
\end{fmfgraph*}
\end{fmffile}
\end{center}
}
+
\parbox{40mm}{
\begin{center}
\begin{fmffile}{zzz-schmematic4}
\begin{fmfgraph*}(30, 8)
\fmfstraight 
\fmfleftn{l}{2}\fmfrightn{r}{2}
\fmfset{arrow_len}{2.8mm}
\fmfset{dash_len}{2.5mm}
\fmf{plain_arrow, tension=2,label=$|n\rangle$,label.side=left, width=1}{r1,m2}
\fmf{plain_arrow, tension=1,label=$|m\rangle$,label.side=left, width=1}{m2,m1}
\fmf{plain_arrow, tension=2,label=$|n\rangle$,label.side=left, width=1}{m1,l1}
\fmffreeze
\fmf{dashes_arrow, left=1}{m1,m2}
\fmfdot{m1}
\fmfdot{m2}
\fmffreeze
\end{fmfgraph*}
\end{fmffile}
\end{center}
}
\right) \ \ \right] 
\\ \nonumber
\end{eqnarray}
where it is sufficient to use the bare Green's function for the defect lines here. It is easy to see $\Gamma_m$ is given by
\begin{align}
\Gamma_n &\sim \sum_{m\, (E_m<E_n)} \int d^3 \mathbf{k} \ [1+\bar{f}(vk/T)] \ U^2 \ \delta(E_n- vk - E_m) + \sum_{m\, (E_m>E_n)} \int d^3 \mathbf{k} \ \bar{f}(vk/T) \ U^2 \ \delta(E_n+vk-E_m) \nonumber \\[.2cm]
& \sim U^2 \epsilon_n^2, \ \ \ \ \ \epsilon_n \sim \mathrm{max}(E_n-E_0, T)
\label{eq:Gamma_n}
\end{align}
where $U$ is the dominating interaction vertex, $\mathbf{k}$ is the loop phonon momentum with $\omega= vk$ its energy, and $\epsilon_n^2$ arises from the density of states. When the first term dominates in the estimation of $\epsilon_n$, the defect state decay through emitting a phonon that is bounded by energy $E_n-E_0$; while when the second term dominates, the defect state absorb a phonon to transit to a higher energy level and the typical phonon energy is of temperature $T$. This gives the estimation of the related phonon energy $\epsilon_n$ and the related density of state factor. For typical models the leading coupling of the phonon to the defect appears as $g\partial u \sim g\sqrt{k}(a+a^\dagger)$, therefore $U\sim g\sqrt{k}$. For a defect that is in resonance with thermal phonons of energy $vk\sim T$, we have $E_n-E_0\sim E-E_0 \sim T$, and therefore $\Gamma_n \sim g^2 T^3$.

Now we can readily compare the contributions of non-resonant processes and resonant processes to the averaged non-skew scattering rate among defects with certain distribution. Consider a phonon with energy $vk\sim T$, its averaged scattering rate is
\begin{align}
\tau_e^{-1}(vk\sim T) &\sim \int d\epsilon \ n(\epsilon) \ \int d^3 \mathbf{k'} \ \delta(vk-vk') \ \left|T^{fi}_{l'l} \right|^2 \nonumber \\[.2cm] 
&\sim \int d\epsilon \ n(\epsilon) \ T^2 \ U_{in} U_{out} U_{in'} U_{out'} G_n G_{n'}
\end{align}
where $\int d\epsilon \: n(\epsilon)$ schematically represents an average over some random ensemble of possible defects. If there is only non-resonant scattering, the typical values of $G$ and $\epsilon$ do not scale with $T$, and there is no involvement of $\Gamma$, so the average scattering rate would scale as
\begin{align}
\text{No resonance:}\quad \tau_e^{-1}(vk\sim T) \sim T^2 U^4 \sim g^4 T^4.
\end{align}
On the other hand, if resonance is available, such that $G \sim \Gamma^{-1} \sim U^{-2} \epsilon^{-2}$ for both $G_n$ and $G_{n'}$, for a narrow range of $\epsilon$ with $|\epsilon-vk| \sim \Gamma$, then the contribution to $W^e$ scales with $U^2$ and hence dominates over the non-resonant $U^4$ (in the end, we are in the regime of perturbation theory). This estimation is indeed valid for resonant non-skew scattering. Therefore, when resonance is available, we have,
\begin{align}
\text{Resonance:}\quad \tau_e^{-1}(vk\sim T) &\sim \Gamma \ n(T) \ T^2 \ U^4 \ \Gamma^{-2} \ \sim \ n(T) \ T^2 \ U^4 \ \Gamma^{-1} \nonumber \\[.2cm]
&\sim n(T) \ U^2 \sim n(T) \ g^2 T.
\label{typical_rate_estimation}
\end{align}
The longitudinal thermal conductivity, dominated by resonant processes, would then scale as
\begin{align}
\kappa_L \sim \int dk \ k^2 \ \bar{f}(vk/T) \ \tau_e(vk) \sim \frac{T^3}{n(T) \: U^2} \sim \frac{T^2}{n(T) \: g^2} \ .
\end{align}
When $n(T)\sim const.$, this is the celebrated result in glasses \cite{Anderson:1972,Phillips:1972}.

Now we consider skew scattering. In the absence of resonance, the estimation is similar to the non-skew, non-resonant case -- except usually one of the four $U$'s must not be the dominating one, but some subleading one $U'\sim g' \omega\sqrt{k}$, where $g'$ is some time-reversal odd parameter, and the $\omega\sim \partial_t$ witnesses the time-reversal oddness. We find 
\begin{align}
\text{No resonance:}\quad \tau_o^{-1}(vk\sim T) \sim T^2 U^3 U' \sim g^3 g' T^5.
\end{align}
The resonant skew scattering case requires some extra care. Again some sub-leading $U'$ can be a source of time-reversal breaking, in which case the skew scattering rate will again be $g'\omega/g$ times the non-skew one. But there is a more dominant contribution. When $G_n$ and $G_{n'}$ are both at resonance, even when the four interactions vertices are all time-reversal even $U$, as long as time-reversal symmetry is broken between $|n\rangle$ and $|n'\rangle$, they may have a small difference in energy $\Delta < \Gamma$, such that the interference makes skew scattering possible. In this case $G_n G_{n'} \sim \Gamma^{-2}$ must not be the valid estimation in $|T^{fi}_{l'l}|^2$, because it does not vanish when the energy difference $\Delta\rightarrow 0$; instead, it is not hard to envision that the valid estimation becomes $G_n G_{n'} \rightarrow G_n\: \Delta \partial_{E_{n'}} G_{n'} \sim \Delta \Gamma^{-3}$. The dominating contribution to resonant skew scattering is therefore (we assume $\Delta$ might be correlated with $\epsilon$, i.e. $\Delta=\Delta(\epsilon)$)
\begin{align}
\text{Resonance:}\quad \tau_o^{-1}(vk\sim T) & \sim \Gamma \ n(T) \ T^2 \ U^4 \ \Delta(T) \Gamma^{-3} \ \sim \ n(T) \ T^2 \ U^4 \ \Delta(T) \Gamma^{-2} \nonumber \\[.2cm]
& \sim n(T) \ \Delta(T) \ T^{-2}.
\end{align}
We emphasize again that this is valid when the small split $\Delta(T)$ is parametrically smaller than the resonance width $\Gamma \sim g^2 T^3$, so even though is does not contain a positive power of $g$, the smallness is implied in $\Delta$. 

Our caricature model in the main text is an example of such scaling, with $\epsilon= 2t - t'$, and $\Delta = \Delta_B \sim B t$. We assumed $n(\epsilon=2t-t')\sim const.$ in the range of temperatures of interest, while $t$ itself does not vary too much and therefore $\Delta \sim const. B$. Since the magnetic flux per unit cell is very small, in the range of temperatures of interest $\Delta < g^2 T^3$ stays valid under reasonable parameters of the system.

To relate the scaling of $\tau_o^{-1}$ to that of $\kappa_H$, we recall from the main text that $\kappa_H/\kappa_L \sim \gamma W^o/W^e$ where $\gamma$ is a dimensionless parameter for typical rotation and inversion breaking effects by \emph{local} scatterers. If $\gamma$ does not scale much with energy, and if both the non-skew and skew scattering are dominated by the same resonating scatterers distributing with $n(\epsilon)$, then we may estimate
\begin{align}
\kappa_H \sim \kappa_L \ \frac{\tau_o^{-1}(T)}{\tau_e^{-1}(T)} \sim \kappa_L \frac{\Delta(T)}{\Gamma} \sim \frac{n(T) \: \Delta(T)}{g^4 \: T} \ \ \ \ \ \left( \ \Delta(T)< \Gamma \sim g^2 T^3 \ \right) \ .
\end{align}
Note that when $n(T)$ and $\Delta(T)$ have low power (or zeroth power) scaling with temperature, $\kappa_H$ can have a low or even negative power scaling, until the temperature becomes so low that $\Delta(T)< \Gamma \sim g^2 T^3$ is no longer valid. If $\gamma$ scales significantly with energy or if the non-skew and skew scatterers are not dominated by the same defects, then the temperature scaling of $\kappa_H$ would depend on more details of the scattering processes.

We remark that there is yet another ``semi-resonance'' scenario to be considered. Suppose resonance scattering is available for non-skew scattering; but for skew scattering, only one of $G_n$ and $G_{n'}$ is at resonance while the other is not. Such scenario would occur, for instance, when the dynamical defect has only two levels, so that while one of $|n\rangle$ and $|n'\rangle$ is the resonating excited state, the other must be the ground state that is non-resonating. In such scenario, the two Green's functions are given by $\Gamma^{-1}$ and $\epsilon^{-1}\sim T^{-1}$ respectively, and hence the semi-resonance contribution to the skew scattering is 
\begin{align}
\text{Semi-resonance:}\quad \tau_o^{-1}(vk\sim T) \sim n(T) \ T^2 \ U^3 U' \ \Gamma^{-1} T^{-1} \sim n(T) T U^3 U' \sim n(T) g^3 g' T^4
\label{semi-res}
\end{align}
which is only one power lower than the non-resonant scattering if $n(T)\sim const.$. The present analysis shows that the full resonance mechanism in our main text dominates over this alternative possibility at low $T$.

\section{Resonance between Nearly Degenerate States}
\label{app_nearly_deg}

In the main text, we considered the time-reversal pair of intermediate states $|\pm\rangle$ to be exactly degenerate to begin with, and they only split upon introducing the magnetic field. In our model this degeneracy is guaranteed by having both the time-reversal symmetry and the $C_4$ symmetry. A natural question to ask is, if in the vicinity of the defect there is some random electric field so that the four sites are subjected to slight different potentials, thus breaking the $C_4$ symmetry and the degenerate pair splits into the time-reversal even and odd combinations $|+\rangle \pm |-\rangle$ with some energy difference $\Delta_E$, how small must $\Delta_E$ be in order for our resonant mechanism to retain. Should it be the more stringent  $|\Delta_E|<|\Delta_B|$ (with $|\Delta_B|<\Gamma$ understood), or should it be just the resonant condition $|\Delta_E|<\Gamma$, regardless of the comparison with $|\Delta_B|$?  In this section we show the latter is valid; our argument is general and does not depend on the particular model. 

In spirit of degenerate perturbation theory, it suffices to project the effects of $\Delta_E$ and $\Delta_B$ onto the subspace of the otherwise degenerate time-reversal pair. In the $|\pm\rangle$ basis, we have the splitting Hamiltonian
\begin{equation}
    \delta H = -\frac{\Delta_E}{2} \sigma^1 - \frac{\Delta_B}{2} \sigma^3
\end{equation}
with eigenvalues $\mp \Delta/2$, $\Delta\equiv \sqrt{\Delta_E^2+\Delta_B^2}$; the new eigenstates are, respectively,
\begin{equation}
\begin{split}
    &|1\rangle \equiv \cos\frac{\alpha}{2} |+\rangle + \sin\frac{\alpha}{2} |-\rangle, \ \ \ \ \ 
    |2\rangle \equiv -\sin\frac{\alpha}{2} |+\rangle + \cos\frac{\alpha}{2} |-\rangle, \\[.2cm]
    &\cos\alpha \equiv \frac{\Delta_B}{\Delta}, \ \ \ \ \ \sin\alpha \equiv \frac{\Delta_E}{\Delta}.
\end{split}
\end{equation}
Now the T-matrix reads
\begin{equation}
    T_{l'l} =\sum_{n=1, 2}U_{n,l'}^{*}\frac{1}{\omega-E_n+i\Gamma_n/2}U_{n,l}
    =\frac{1}{2}\frac{2(\omega-\epsilon+i\Gamma/2) \Omega^0_{l'l} - \Delta_E \Omega^E_{l'l} - \Delta_B \Omega^B_{l'l}}{(\omega-(\epsilon-\Delta/2)+i\Gamma/2)(\omega-(\epsilon+\Delta/2)+i\Gamma/2)}
\end{equation}
where we defined
\begin{equation}
\begin{split}
    &\Omega^0_{l'l} \equiv U^\dagger_{l'} U_l \equiv U^*_{+,l'}U_{+,l} + U^*_{-,l'}U_{-,l}, \\
    &\Omega^E_{l'l} \equiv U^\dagger_{l'} \sigma^1 U_l \equiv U^*_{+,l'}U_{-,l} + U^*_{-,l'}U_{+,l}, \\ 
    &\Omega^B_{l'l} \equiv U^\dagger_{l'} \sigma^3 U_l \equiv U^*_{+,l'}U_{+,l} - U^*_{-,l'}U_{-,l}.
\end{split}
\end{equation}
For skew scattering, we are only interested in the terms in $|T_{l'l}|^2$ that are odd in $\Delta_B$. (There may be extra $B$ dependence in the couplings $U_{\pm, l}$ to begin with. However, by symmetry considerations, the effects from those couplings would not receive full resonance enhancement, and therefore we will ignore them. One such term would be the Lorentz force term $\sum_i \bB\cdot(e^\star \mathbf{R}_i c^\dagger_{\mathbf{R}_i} c_{\mathbf{R}_i} \times \dot\bu)$ in our model of ionic defect.)
We need the following properties of the $\Omega$'s:
\begin{equation}
\begin{split}
    &\Omega^B_{l'l} \Omega^{0*}_{l'l} + c.c. = 2|U_{+,l'}|^2|U_{+,l}|^2-2|U_{-,l'}|^2|U_{-,l}|^2 = 0, \\[.2cm]
    &\Omega^B_{l'l} \Omega^{E*}_{l'l}+ c.c. = \left( |U_{+,l'}|^2-|U_{-,l'}|^2\right) \left(U_{+,l}U^*_{-,l} +  U^*_{+,l}U_{-,l}\right) + \left( |U_{+,l}|^2-|U_{-,l}|^2\right)\left(U^*_{+,l'}U_{-,l'} +  U_{+,l'}U^\star_{-,l}\right) = 0
\end{split}
\end{equation}
where we have used the time-reversal symmetry between the $|\pm\rangle$ states together with the $\bk\leftrightarrow -\bk$ inversion symmetry of the skew scatterers, which we presumed throughout this paper. Thus, in the time-reversal odd part of $|T_{l'l}|^2$, the only appearance of $\Delta_E$ is to modify the $\mp\Delta_B/2$ in the denominator into $\mp\Delta/2$; as long as the resonance condition $|\Delta|<\Gamma$ remains satisfied, the effect of having a non-zero $\Delta_E$ is of order $1$ compared to when $\Delta_E=0$.

The arguments so far apply to the resonant skew scattering due to a single defect. In the main text we also mentioned an alternative scenario of interference between two nearby identical defects, following Ref.~\cite{Mori:2014}. In this scenario, suppose the two defects are subjected to the same $\Delta_B$, but the random electric potentials might be slightly different, so that they have different $\Delta_E$ and $\Delta_E'$, and moreover, slightly different $\epsilon$ and $\epsilon'$ whose difference is of order $\Delta_E$. The two defect interference contribution to skew scattering is proportional to
\begin{equation}
    e^{-i(\bk'-\bk)\cdot\delta\br} T'^*_{l'l} T_{l'l} + e^{i(\bk'-\bk)\cdot\delta\br} T^*_{l'l} T'_{l'l}.
\end{equation}
There are two differences compared to the case without the small random electric field. The first difference is that the four energies in the denominator are now $\epsilon\mp\Delta$ and $\epsilon'\mp\Delta'$; but again, the effect of this difference is an order $1$ factor as long as the resonance condition remains satisfied, i.e. $|\Delta|, |\Delta'|, |\epsilon'-\epsilon|$ remaining smaller than $\Gamma$. The second difference is that, in the numerator, in addition to the original $\Delta_B\Gamma \cos[(\bk'-\bk)\cdot\delta\br]$ term, there will also be new terms that involve $\Delta_B (\epsilon'-\epsilon)\sin[(\bk'-\bk)\cdot\delta\br]$ and $\Delta_B (\Delta_E'-\Delta_E)\sin[(\bk'-\bk)\cdot\delta\br]$. However, these new terms are odd in $\delta\br$; as long as the random electric field is uncorrelated with $\delta\br$, the effects of these new terms vanish under the $\delta\br$ integral with the distribution $p(\delta\br)$ which is even by definition. If the random electric field is somewhat weakly correlated with $\delta\br$, these new terms will give some contribution, but smaller than the original one, due to the resonance condition.

\section{Inverting the Collision Kernel}
\label{app_invert}

In this appendix we include the details of inverting the Boltzmann collision kernel. It is convenient to decompose the angular dependence of the collision kernel into spherical harmonics $Y_j^{j_z}(\theta_\bk, \phi_\bk)$. For simplicity we will ignore the dependence on the polarization $\alpha, \alpha'$ below; including such dependence only makes order $1$ differences. Suppose the collision probability takes the form
\begin{align}
    W^e_{\bk'\bk}=& \ {\frac{\delta(\omega_{\bk'}-\omega_\bk)}{V}} \frac{4\pi}{\mathcal{D}(\omega_\bk)} \left\{ \ \ w^e_0(\omega_\bk) \ Y'{}_0^0 Y_0^0 + w^e_1(\omega_\bk) \sum_{j_z=\pm 1, 0} Y'{}_1^{j_z} Y_1^{-j_z} + w^e_2(\omega_\bk) \sum_{j_z=\pm 2, \pm 1, 0} Y'{}_2^{j_z} Y_2^{-j_z}
    \right. \nonumber \\[.1cm]
    & \hspace{2.35cm} \left. + \: w^e_{1+2}(\omega_\bk) \left( e^{-i3\phi_0} Y'{}_1^{+1} Y_2^{+2} + e^{+i3\phi_0} Y'{}_1^{-1} Y_2^{-2} - e^{-i3\phi_0} Y'{}_2^{+2} Y_1^{+1} - e^{+i3\phi_0} Y'{}_2^{-2} Y_1^{-1} \right) \right. \nonumber \\[.1cm]
    & \hspace{2.35cm} \left. + \: w^e_{1-2}(\omega_\bk) \left( e^{+i\psi_0} Y'{}_1^{+1} Y_2^{-2} + e^{-i\psi_0} Y'{}_1^{-1} Y_1^{+2} - e^{+i\psi_0} Y'{}_2^{+2} Y_1^{-1} - e^{-i\psi_0} Y'{}_2^{-2} Y_1^{+1} \right) \phantom{\sum_1^1 \!\!\!\!} \right\} \: , \\[.2cm]
    W^o_{\bk'\bk} =& \ {\frac{\delta(\omega_{\bk'}-\omega_\bk)}{V}} \: \frac{4\pi}{\mathcal{D}(\omega_\bk)} \: w^o(\omega_\bk) \left( iY'{}_2^{+2}Y_2^{-2} - iY'{}_2^{-2} Y_2^{+2} \right) 
\end{align}
where $\mathcal{D}\equiv 4\pi\omega_\bk^2/(2\pi v)^3$ is the density of states, and $Y$ and $Y'$ have arguments $(\theta_\bk, \phi_\bk)$ and $(\theta_{\bk'}, \phi_{\bk'})$ respectively. The first line of $W^e$ are the rotational invariant terms with $j=0, 1, 2$ respectively; higher $j$ contributions will be irrelevant to our discussion. The second and third lines of $W^e$ are the rotation and inversion symmetry breaking terms that transit between $j=1$ and $j=2$, as mentioned in the main text; we only kept the $j_z=\pm j$ parts that will be relevant. The $W^o$ is the simplest resonance skew scattering compatible with the general discussions that led to Eq.~\eqref{res_skew_k_dependence}, and is indeed the form that would appear from our caricature model.

The basic idea here is that, according to our general discussion of Boltzmann equation and in particular Eq.~\eqref{kappaH_expression} in the main text, the components in $(C^e)^{-1}$ and $[(C^e)^{-1} C^o (C^e)^{-1}]$ that will give rise to $\kappa_L$ and $\kappa_H$ are those with $j=|j_z|=1$. Meanwhile, $W^o$ only involves $j=|j_z|=2$. It is the $w^e_{1\pm 2}$ terms that connect between these two sets of angular momentum components. 
Let us therefore write down the matrix elements of the collision kernel $C$ that involve $j=1, j_z=\pm 1$ and $j=2, j_z=\pm 2$. More particularly, define $\wt{Y}_1^{\pm1} \equiv e^{\mp i(3\phi_0-\psi_0)/2} Y_1^{\pm 1}$ and $\wt{Y}_2^{\pm 2} \equiv e^{\mp i(3\phi_0+\psi_0)/2} Y_2^{\pm 2}$, then in the basis of $(\wt{Y}_1^{+1}+\wt{Y}_1^{-1})/\sqrt{2}$, $(-i\wt{Y}_1^{+1}+i\wt{Y}_1^{-1})/\sqrt{2}$, $(\wt{Y}_2^{+2}+\wt{Y}_2^{-2})/\sqrt{2}$, $(-i\wt{Y}_2^{+2}+i\wt{Y}_2^{-2})/\sqrt{2}$, we have
\begin{align}
\hat{C}^e = \: -w^e_0 \: + \left[ \begin{array}{cccc} w^e_1 & & & \\[.1cm] & w^e_1 & & \\[.1cm] & & w^e_2 & \\[.1cm] & & & w^e_2 \end{array} \right] + (1+2\bar{f}) \left[ \begin{array}{cccc} & & w^e_{1-2} + w^e_{1+2} &  \\[.1cm] & & & w^e_{1-2}-w^e_{1+2} \\[.1cm] -w^e_{1-2}-w^e_{1+2} & & & \\[.1cm] & -w^e_{1-2}+w^e_{1+2} & & \end{array} \right] \ ,
\end{align}
\begin{align}
\hat{C}^o = (1+2\bar{f})\left[ \begin{array}{cccc} & & & \\[.1cm] & & & \\[.1cm] & & & w^o \\[.1cm] & & -w^o & \end{array} \right] \ ,
\end{align}
where $C=\hat{C} (\delta(\omega_{\bk'}-\omega_\bk)/V)(4\pi/\mathcal{D}(\omega_\bk))$. Now we invert the collision kernel. We will only care about the $j=1$ block of the inversion, since according to Eq.~\eqref{kappaH_expression} only the $j=1$ block will contribute to $\kappa_L$ and $\kappa_H$. We find
\begin{align}
    \left(\hat{C}^e\right)^{-1}_{j=1} = \left[ \begin{array}{cc} \frac{-w^e_0+w^e_2}{\det_+} & \\[.1cm] & \frac{-w^e_0+w^e_2}{\det_-} \end{array} \right] \ ,
\end{align}
\begin{align}
    -\left[\left(\hat{C}^e\right)^{-1} \hat{C}^o \left(\hat{C}^e\right)^{-1} \right]_{j=1} = \frac{(1+2\bar{f})^3 \left[ (w^e_{1+2})^2 - (w^e_{1-2})^2 \right] \: w^o }{\det_+ \det_-} \left[ \begin{array}{cc}  & -1 \\[.1cm] 1 & \end{array} \right] \ ,
\end{align}
where $\det_\pm \equiv (w^e_0-w^e_1)(w^e_0-w^e_2)+(1+2\bar{f})^2(w^e_{1-2} \pm w^e_{1+2})^2$. 

Note that the matrix $(\hat{C}^e)^{-1}_{j=1}$ is not rotational invariant since $\det_+\neq \det_-$ in general. As our basis depends on the local defect orientations $\phi_0$ and $\psi_0$,  upon random averaging them (see main text for physical discussion), only the rotational invariant part proportional to the identity matrix remains:
\begin{align}
    \left(\hat{C}^e\right)^{-1}_{j=1} \longrightarrow \ \left(-w^e_0+w^e_2\right)\frac{\det_+ + \det_+}{2\det_+ \det_-} \mathbf{1}_{2\times 2} \equiv -\tau(\omega_\bk) \mathbf{1}_{2\times 2}
\end{align}
where $\tau$ is defined to be the relaxation time extracted from the longitudinal thermal transport. In particular, from Eq.~\eqref{kappaH_expression} we can read-off
\begin{align}
    \kappa_L = \int d\omega \, \mathcal{D} \, \frac{v^2}{3} \omega \, \tau \, \frac{\partial\bar{f}}{\partial T}
\end{align}
(recall $\hat{\bf x}=\sqrt{4\pi/3} (Y_1^{+1}+Y_1^{-1})/\sqrt{2}$, $\hat{\bf y}=\sqrt{4\pi/3} (-iY_1^{+1}+ iY_1^{-1})/\sqrt{2}$). In the usual relaxation time approximation, only $w^e_0$ is kept non-zero, in which case $\tau=1/w^e_0$.

On the other hand, the matrix $[(C^e)^{-1} C^o (C^e)^{-1}]_{j=1}$ is rotational invariant to begin with, and hence its effect is manifestly independent of $\phi_0$ and $\psi_0$. From Eq.~\eqref{kappaH_expression} we can read-off
\begin{align}
    \kappa_H = \int d\omega \, \mathcal{D} \, \frac{v^2}{3} \omega \, \left(\gamma w^o\tau\right) \tau \left( 1+2\bar{f}\right)^3 \frac{\partial\bar{f}}{\partial T} \ ,
    \label{eq:kappa_H_app}
\end{align}
(and we can denote $\tau_o=1/w^o$) where we defined the dimensionless ratio
\begin{align}
    \gamma \equiv \frac{(w^e_{1+2})^2 - (w^e_{1-2})^2 }{\tau^2 \ \det_+ \det_-} \ .
\end{align}
If $w^e_0$ dominates over all other components, then $\gamma \simeq [(w^e_{1+2})^2 - (w^e_{1-2})^2]/(w^e_0)^2$, so we may think of $\gamma$ as some kind fractional local (co-)variance of the rotation and inversion breaking terms, as said in the main text.

\section{Details of the Three-level Model}
\label{app_model_3-level}

\subsection{Microscopic Origin of the Coupling between Defect and Acoustic Motion}
\label{app_g_origin}
To make more physical sense of the coupling $g$ we think of the transition of states in the four sites as the process of moving defect ions around four possible potential wells. The potential well location depends on the acoustic displacement $u$ in its vicinity, as described by the potential $V(x-u)$. We can write down the kinetic energy of the defect ion and its interaction with the crystal environment as follows:
\begin{equation}
    p^2/2m_{def}+ V(x-u)+ P^2/2\bar{M}_{ion}+ V_{ion}(u)+....
\end{equation}
where $\bar{M}_{ion}$ is some suitable average of environment ions' masses. (In the main text $\Pi$ is the momentum density conjugate to $u$, while here for illustrative purpose it is more convenient to use $P$, the momentum conjugate to $u$.) We can now do a canonical transformation which preserves the commutation relations:
\begin{equation}
\begin{split}
    &x'= x-u, \quad p'= p,\\
    &u'= u, \quad P'= P+p.
\end{split}
\end{equation}
In this new basis, the defect ion and phonons couple through $(p'\cdot P')/\bar{M}_{ion}$ term. For a defect moving on a four site square of size $b$, the momentum $p'$ takes typical value of $2\pi/b$. In continuum, this interaction term becomes the coupling $g$ in our main text with $g$ being order $1/b$.

\subsection{Three-level Caricature Model}
\label{app_calculations}

In this subsection, we present the detailed calculation for the phonon scattering rate at finite temperature from the three-level defect model defined in the main text. The free phonon field is quantized as
\begin{equation}
    u_{i}(\bx)=\sum_{\alpha\bk} \frac{1}{\sqrt{2\rho\omega_{\alpha \bk}V}}e^{\alpha}_{i}(\bk)(a_{\alpha\bk}e^{i\bk\cdot\bx}+a_{\alpha\bk}^{\dagger}e^{-i\bk\cdot\bx});
\end{equation}
\begin{equation}
    \Pi_{i}(\bx)=-i \sum_{\alpha\bk}\sqrt{\frac{\rho\omega_{\alpha\bk}}{2 V}}e^{\alpha}_{i}(\bk)(a_{\alpha\bk}e^{i\bk\cdot\bx}-a_{\alpha\bk}^{\dagger}e^{-i\bk\cdot\bx}),
\end{equation}
where the polarization vectors satisfies the normalization condition:
\begin{equation}
    \sum_{\alpha}e^{\alpha}_{i}e^{\alpha}_{j}=\delta_{ij},\quad \sum_{i}e_{i}^{\alpha}e_{i}^{\beta}=\delta_{\alpha\beta},
\end{equation}
The canonical commutation relation reads
\begin{equation}
[u_{i}(\bx),\Pi_{j}(\bx')]=i\delta^3(\bx-\bx')\delta_{ij},\quad [a_{\alpha\bk},a_{\alpha'\bk'}^{\dagger}]=\delta_{\alpha\alpha'}\delta_{\bk\bk'}.
\end{equation}
The related matrix element of defect-phonon interaction can be worked out
\begin{equation}
\begin{split}
    &\langle +|H_{int}|0,\alpha\bk\rangle=-i\frac{\sqrt{\omega_{\alpha\bk}/\rho V}}{2}g\left( e_{x}^{\alpha}-ie_{y}^{\alpha} \right);\\
    &\langle -|H_{int}|0,\alpha\bk\rangle=i\frac{\sqrt{\omega_{\alpha\bk}/\rho V}}{2}g\left( e_{x}^{\alpha}+ie_{y}^{\alpha} \right);
\end{split}
\end{equation}
where $|\pm\rangle$ and $|0\rangle$ are the three states we are interested in corresponding to energy $E_{\pm}$ and $E_0$, respectively; $\alpha \bk$ labels the polarization and momentum of the phonons.

Now we can compute the T-matrix for the resonant scattering. In doing so we need the lifetime of each possible intermediate state.  In our model, we consider a singlet $|0\rangle$ and a doublet $|\pm\rangle$ with an energy difference of $\epsilon$, where $\epsilon>0$ represents a higher energy for the doublet and we will assume a random distribution of $\epsilon$ for the defects which also include cases of $\epsilon<0$. At finite temperature, all three states can obtain a finite life time which to lowest order in the magnetic field can be written as:
\begin{equation}
\begin{split}
    \Gamma(T,\epsilon)\equiv\Gamma_{+}^{(0)}=\Gamma_{-}^{(0)}=\text{sgn}(\epsilon)\frac{e^{\beta \epsilon}}{e^{\beta \epsilon}-1}\Gamma(0,\epsilon), \ \ \ \ \ \Gamma_0(T,\epsilon)=\text{sgn}(\epsilon)\frac{1}{e^{\beta\epsilon}-1}2\Gamma(0,\epsilon),
\end{split}
\end{equation}
where the $1/\Gamma(0,\epsilon)$ is the lifetime of the doublet as an excited state at zero temperature and $\beta\equiv \frac{1}{T}$ is the inverse temperature. The factor of $2$ comes from two ways of absorbing resonant thermal phonons of energy $\epsilon$. According to Fermi's golden rule (i.e. at the leading 1-loop order, see Appendix \ref{app_formal}), at zero temperature we have
\begin{equation}
    \Gamma(0,\epsilon)=\frac{\pi}{3}\left( \frac{|\epsilon| g^2}{\rho} \right) D(|\epsilon|)=\frac{|\epsilon|^3 g^2}{2\pi\rho v^3},
\end{equation}
where $D(\epsilon)$ is the density of states of the phonon at energy $\epsilon$ including all polarizations and $v$ is a proper average of the sound speed in different polarizations, $1/v^3\equiv(1/v_L^3+2/v_T^3)/3$. Here we have used the angular averaged matrix element modulus squared, given by $\epsilon g^2/6\rho$. Therefore
\begin{equation}
\begin{split}
    \Gamma(T,\epsilon)=\frac{e^{\beta \epsilon}}{e^{\beta \epsilon}-1}\frac{\epsilon^3 g^2}{2\pi\rho v^3}, \ \ \ \ \ \Gamma_0(T,\epsilon)=\frac{1}{e^{\beta\epsilon}-1} \frac{\epsilon^3 g^2}{\pi \rho v^3}= 2e^{-\beta\epsilon}\Gamma(T,\epsilon).
\end{split}
\end{equation}
(In fact we will not use $\Gamma_0$ below, only $\Gamma$.)

Now for the T-matrix, we can separate the $|T_{l'l}^{fi}|^2$ ($f,i\in{0,\pm}$ are the final and initial states of the defect) into symmetric (non-skew scatterings) and antisymmetric part (skew scatterings). We will focus on the skew scattering part here. The resonant skew scattering can only occur when the intermediate states of the defect is the doublet such that there are two scattering processes with the same initial and final states to interfere. For $\epsilon>0$ ($\epsilon<0$), the resonant process includes the defect virtually obsorbing (emitting) the incoming (outgoing) phonon and then emitting (obsorbing) an outgoing (incoming) phonon of energy $|\epsilon|$. So only in $|T_{l'l}^{00}|^2$ there exist antisymmetric components and this T- matrix can be written as:
\begin{equation}
 T_{l'l}^{00}=U_{+,l'}^{*}\frac{1}{\pm\omega-(\epsilon-\Delta_B/2)+i\Gamma(T,\epsilon)/2}U_{+,l}+U_{-,l'}^{*}\frac{1}{\pm\omega-(\epsilon+\Delta_B/2)+i\Gamma(T,\epsilon)/2}U_{-,l},
\end{equation}
where $U_{\pm l}\equiv \langle \pm |H_{int}|0,\alpha\bk\rangle$ and the $\pm$ represent two cases of $\epsilon>0$ or $\epsilon<0$. For both cases, the anti-symmetric part of its modulus square, to leading order in $\phi$, is
\begin{equation}
\begin{split}
    |T_{l'l}^{00}|^2_{A}
    &= i\frac{\Gamma(T,\epsilon)\Delta_B}{2\left[(\omega-|\epsilon|)^2+\Gamma(T,\epsilon)^2/4\right]^2}(U_{+,l'}^* U_{+,l} U_{-,l}^* U_{-,l'}-U_{+,l}^*U_{+,l'}U_{-,l'}^{*}U_{-,l})\qquad \text{up to linear order in}\;\phi,\\
    &=\frac{ g^4\omega^2  \Gamma(T,\epsilon)\Delta_B}{16\left[(\omega-|\epsilon|)^2+\Gamma(T,\epsilon)^2/4\right]^2 \rho^2V^2}F_{\alpha'\alpha}(\hat{\bk}',\hat{\bk}),
\end{split}
\end{equation}
where apart from the photon energy dependent factor, the matrix $F$ contains the information about the polarization $\alpha$, $\alpha'$ and directions of the incoming and outgoing phonon momenta $\hat{\bk}$, $\hat{\bk}'$. In particular, the phonon energy dependent part is the important factor that we will use to study the temperature scaling of thermal conductivity tensor. For later estimation purpose, we also write down the explicit formula for $F_{\alpha'\alpha}(\hat{\bk}',\hat{\bk})$:
\begin{equation}
    F=\sin^2(\theta_{\alpha\hat{\bk}})\sin^2(\theta_{\alpha'\hat{\bk}'})\sin 2(\phi_{\alpha\hat{\bk}}-\phi_{\alpha'\hat{\bk}'}),
\end{equation}
where $(\theta_{\alpha\hat{\bk}},\phi_{\alpha\hat{\bk}})$ are the inclination and azimuth angles for the polarization vector $\mathbf{e}^{\alpha}$ at momentum along $\hat{\bk}$ direction. For the longitudinal mode, $\mathbf{e}^{\mathrm{long.}}=\hat\bk$, this angular factor is indeed proportional to the  $iY'{}_2^{+2}Y_2^{-2}-iY'{}_2^{-2}Y_2^{+2}$ that we introduced for $W^o$ in the previous appendix; for the transverse modes, the identification is understood with a rotation from $\hat{\bk}$ to the polarization direction $\mathbf{e}^{\alpha}$. (Notice that $\sum_{\alpha',\alpha}F_{\alpha'\alpha}(\hat{\bk'},\hat{\bk})=0$. Nonetheless, this equal weight sum will not arise in the calculation for $\kappa_{L, H}$ -- instead, such sum will always appear with coefficients that are different by order $1$ ratios and hence no significant cancellation would occur, because the longitudinal and transverse phonon velocities differ by order $1$.)

For three-level defects of a distribution $n(\epsilon)$ (per energy per volume) of level spacing $\epsilon$, the skew scattering rate to leading order in the magnetic field is
\begin{equation}
\begin{split}
W_{l'l}^o&=\int d\epsilon \ n(\epsilon)\frac{2\pi}{V}\delta(\omega_l-\omega_l')\frac{1}{1+2e^{-\beta\epsilon}}\frac{ g^4\omega_l^2  \Gamma(T,\epsilon)\Delta_B}{16\rho^2\left[(\omega_l-|\epsilon|)^2+\Gamma(T,\epsilon)^2/4\right]^2}F_{\alpha'\alpha}(\hat{\bk}',\hat{\bk})\\
&\approx \frac{\pi^2}{2 V}\left[n(\omega_l)\frac{1}{1+2e^{-\beta\omega_l}}\frac{g^4 \omega_l^2\Delta_B}{\rho^2\Gamma(T,\omega_l)^2}+n(-\omega_l)\frac{1}{1+2e^{\beta\omega_l}}\frac{g^4 \omega_l^2\Delta_B}{\rho^2\Gamma(T,-\omega_l)^2}\right]\delta(\omega_l-\omega_{l'})F_{\alpha'\alpha}(\hat{\bk}',\hat{\bk}),
\end{split}
\end{equation}
where the $(1+2e^{-\beta\epsilon})^{-1}$ factor is the thermal probability that the initial state is in the $|0\rangle$ state, so to enable a resonance skew scattering. The two terms comes from two resonance possibilities, $\epsilon=\omega>0$ and $\epsilon=-\omega<0$, respectively.
As explained in the main text, the skew scattering rate here has angular distribution of momenta in a way that does not generate thermal Hall effect if the non-skew scattering is isotropic. One possible way to have non-trivial contribution to the thermal Hall effect is to consider other rotational symmetry breaking defects. For the purpose of a simple estimation, we can assume the phonon polarization to be longitudinal. From Eq.~\eqref{eq:kappa_H_app}, we can see that considering other rotational symmetry breaking defects, the full thermal Hall conductivity is:
\begin{align}
    \kappa_H = \int d\omega \, \mathcal{D} \, \frac{v^2}{3} \omega \, \left(\gamma w^o\tau\right) \tau \left( 1+2\bar{f}\right)^3 \frac{\partial\bar{f}}{\partial T} \ ,
\end{align}
and from $W^o$ above we can read off $w^o$ for longitudinal phonons: 
\begin{equation}
\begin{split}
    w^o&=\frac{\pi}{4}\sqrt{\frac{2\pi}{15}}\mathcal{D}\left[n(\omega)\frac{1}{1+2 e^{-\beta\omega}}\frac{g^4\omega^2\Delta_B}{\rho^2\Gamma(T,\omega)^2}+n(-\omega)\frac{1}{1+2 e^{\beta\omega}}\frac{g^4\omega^2\Delta_B}{\rho^2\Gamma(T,-\omega)^2}\right]\\
    &\approx \frac{\pi}{2}\sqrt{\frac{2\pi}{15}}n(0)\Big{[}\frac{1}{1+2 e^{-\beta\omega}}\left(\frac{e^{\beta\omega}-1}{e^{\beta\omega}}\right)^2+\frac{1}{1+2 e^{\beta\omega}}\left(\frac{e^{-\beta\omega}-1}{e^{-\beta\omega}}\right)^2\Big{]}\frac{\Delta_B  v^3}{\omega^2} \ .
\end{split}
\end{equation}
Considering more details such as contribution from other phonon polarizations and a detailed modelling of non-skew scattering part will only give some $\mathcal{O}(1)$ correction. Furthermore, for simplicity, we will assume that the dimensionless parameter $\gamma$ representing the degree of rotational (inversion) symmetry breaking in non-skew scattering has no (or mild) energy and temperature dependence. This will not qualitatively change the later discussion of low temperature scaling if $\gamma$ approaches some constant value at low energies. We also assume that $n(\epsilon)$ has a broad distribution of $\epsilon$ at low energies such that $n(\epsilon)\sim n(0)$, until a soft cutoff $\Lambda$ (e.g. $\Lambda\sim 20\mathrm{K}$) such that when $|\epsilon|>\Lambda$, $n(\epsilon)$ becomes small. On the other hand, there are small energy cutoffs $\lambda_\pm$, such that for $-\lambda_-<\epsilon<\lambda_+$, the resonance condition $|\Delta_B|<\Gamma$ fails; for typical temperatures of interest $\lambda_-\approx\lambda_+$, which we denote as $\lambda$. (In fact, the resonance condition would fail for large negative $\epsilon$ as well, but for typical temperatures of interest this is beyond $-\Lambda$ already.)
The integration of phonon energy, which satisfies $\omega\approx\pm\epsilon$ at resonance, is thus effectively performed in-between, $\lambda<\omega<\Lambda$. We find the low temperature behavior of the thermal Hall conductivity to be
\begin{equation}
    \kappa_H\sim \mathcal{F}(\lambda/T,\Lambda/T) \gamma n(0)\Delta_B T (v\tau)^2,
\end{equation} 
where the $\mathcal{F}$ is some numerical factor that depends on the cutoffs $\lambda$ and $\Lambda$. In particular, the dependence on the infrared cutoff turns out to be logarithmic for small $\lambda/T$, and for $T\sim \Lambda$ (e.g., $T=10\mathrm{K}$, $\Lambda=20\mathrm{K}$) the resulting factor $\mathcal{F}$ is of order 1. For our estimation in the main text, we neglect the dimensionless factor $\mathcal{F}$, phonon polarizations, and other possible microscopic details of the scattering. We can then estimate the thermal Hall conductivity as $\kappa_H\sim \gamma n(0)\Delta_B T(v\tau)^2$. If we take the parameters used in the main text: $t=100 K$, $e^{\star}=-2|e|$, $b=0.1\Ang$, $g=\hbar/b$, $\rho=0.087 \mathrm{K^{-1}\AA^{-5}}$ ($\approx 7000 \mathrm{kg}/\mathrm{m}^3$) and $v=300 \mathrm{K \AA}$ ($\approx 3900 \mathrm{m/s}$), then for a magnetic field of $15$\text{T}, the magnetic length is $l_B=\sqrt{\hbar c/(|e|B)}\approx 6.61\mathrm{nm}$. For $b=0.1\Ang$, the magnetic splitting energy is $\Delta_B=-2t b^2/l_B^2\approx-0.00046 K$, where we have assumed the effective charge $e^{\star}=-2|e|$.

\section{Theoretical Background of the Resonance Computation}
\label{app_formal}

In this last section, we revisit the recipe for computing the resonant scattering, in particular the standard assumptions understood throughout this work:
\begin{enumerate}
\item The resonance width of the dynamical defect is dominated by the one phonon loop process.
\item The collision kernel is dominated by the one phonon in, one phonon out process.
\end{enumerate}
While the recipe is standard and has been commonly used for decades, the theoretical justification behind might have become unfamiliar nowadays, therefore we explain the justification in details in this section.

The calculation recipe is that of  degenerate perturbation theory, with a suitable resummation that captures the leading effects. To justify the use of perturbation theory, we first assume that there is an energy cutoff $\omega_c$ of the theory, where $\omega_c$ should be well above the temperature of interest, while well below the microscopic scale (the Debye frequency in our case) so that we can focus on the relevant degrees on freedom (acoustic phonons near the band bottom in our case); that is, $T<\omega_c<\omega_{Debye}$. Then, in order to view the interactions between the phonons and the defects as perturbations, we need to assume that, for phonons with energies up to the cutoff $\omega_c$, their interaction energy with the defect is small compared to the defect's own energy gap. This justifies the use of perturbation theory. However, the perturbative interactions are acting on a degenerate non-interacting basis, because the phonon spectrum is continuous and may cross the defect's energy gap (i.e. resonance), therefore we need a suitable resummation to properly solve the degenerate perturbation theory.

The full defect Green's function for the $|n\rangle$ state, to 1-loop self-energy correction, is expressed as
\begin{eqnarray}
&& iG_n \  = \ \parbox{20mm}{
\begin{center}
\begin{fmffile}{zzz-full-prop1}
\begin{fmfgraph*}(16, 8)
\fmfstraight 
\fmfleftn{l}{3}\fmfrightn{r}{3}
\fmfset{arrow_len}{2.8mm}
\fmfset{dash_len}{2.5mm}
\fmf{plain_arrow, tension=1,label=$|n\rangle$,label.side=left, width=1.8}{r2,l2}
\fmfdot{r2,l2}
\end{fmfgraph*}
\end{fmffile}
\end{center}
}
\ = \ i\left(G_n^{bare, -1} - \Sigma_n\right)^{-1}, \nonumber \\[.3cm]
&& -i\Sigma_n \ = \ \sum_m\left(\parbox{38mm}{
\begin{center}
\begin{fmffile}{zzz-self-energy1}
\begin{fmfgraph*}(30, 8)
\fmfstraight 
\fmfleftn{l}{2}\fmfrightn{r}{2}
\fmfset{arrow_len}{2.8mm}
\fmfset{dash_len}{2.5mm}
\fmf{plain_arrow, tension=2,label=$|n\rangle$,label.side=left, width=1}{r1,m2}
\fmf{plain_arrow, tension=1,label=$|m\rangle$,label.side=left, width=1}{m2,m1}
\fmf{plain_arrow, tension=2,label=$|n\rangle$,label.side=left, width=1}{m1,l1}
\fmffreeze
\fmf{dashes_arrow,right=1}{m2,m1}
\fmfdot{m1}
\fmfdot{m2}
\fmffreeze
\end{fmfgraph*}
\end{fmffile}
\end{center}
}
+
\parbox{38mm}{
\begin{center}
\begin{fmffile}{zzz-self-energy2}
\begin{fmfgraph*}(30, 8)
\fmfstraight 
\fmfleftn{l}{2}\fmfrightn{r}{2}
\fmfset{arrow_len}{2.8mm}
\fmfset{dash_len}{2.5mm}
\fmf{plain_arrow, tension=2,label=$|n\rangle$,label.side=left, width=1}{r1,m2}
\fmf{plain_arrow, tension=1,label=$|m\rangle$,label.side=left, width=1}{m2,m1}
\fmf{plain_arrow, tension=2,label=$|n\rangle$,label.side=left, width=1}{m1,l1}
\fmffreeze
\fmf{dashes_arrow, left=1}{m1,m2}
\fmfdot{m1}
\fmfdot{m2}
\fmffreeze
\end{fmfgraph*}
\end{fmffile}
\end{center}
}\right) \\[-.1cm] \nonumber
\end{eqnarray}
where the time propagates from right to left. We can view the Feynman diagrams either in terms of old fashioned perturbation theory or in terms of Feynman's time ordered perturbation theory, as the two formalism are manifestly identical in this case because the defect has only a ``particle'' but no corresponding ``hole'' or ``anti-particle'' for pair creation or annihilation. The bare defect Green's function of the $|m\rangle$ state given by
\begin{eqnarray}
iG_m^{bare} \  = \ \parbox{20mm}{
\begin{center}
\begin{fmffile}{zzz-bare-prop1}
\begin{fmfgraph*}(16, 8)
\fmfstraight 
\fmfleftn{l}{3}\fmfrightn{r}{3}
\fmfset{arrow_len}{2.8mm}
\fmfset{dash_len}{2.5mm}
\fmf{plain_arrow, tension=1,label=$|m\rangle$,label.side=left, width=1}{r2,l2}
\fmfdot{r2,l2}
\end{fmfgraph*}
\end{fmffile}
\end{center}
}
\ = \ \frac{i |m\rangle \langle m|}{\omega - E_m + i0^+} \ .
\end{eqnarray}
Here the thermal Boltzmann weight of the defect's initial state is not included in the definition of the Green's function; it will take it into account as an extra classical probability of the initial condition in calculations. On the other hand, the phonon propagators are understood in the thermal ensemble, with the right-to-left line adding a phonon to the thermal ensemble and the left-to-right line removing a phonon from the thermal ensemble:
\begin{eqnarray}
\parbox{20mm}{
\begin{center}
\begin{fmffile}{zzz-phonon-prop1}
\begin{fmfgraph*}(16, 8)
\fmfstraight 
\fmfleftn{l}{3}\fmfrightn{r}{3}
\fmfset{arrow_len}{2.8mm}
\fmfset{dash_len}{2.5mm}
\fmf{dashes_arrow, tension=1,label=$\mathbf{k}\alpha$,label.side=left, width=1}{r2,l2}
\fmfdot{r2,l2}
\end{fmfgraph*}
\end{fmffile}
\end{center}
}=\frac{i (1+\bar{f}(v_\alpha k/T))}{\omega - v_\alpha k + i0^+},
\hspace{2cm}
\parbox{20mm}{
\begin{center}
\begin{fmffile}{zzz-phonon-prop2}
\begin{fmfgraph*}(16, 8)
\fmfstraight 
\fmfleftn{l}{3}\fmfrightn{r}{3}
\fmfset{arrow_len}{2.8mm}
\fmfset{dash_len}{2.5mm}
\fmf{dashes_arrow, tension=1,label=$\mathbf{k} \alpha$,label.side=right, width=1}{l2,r2}
\fmfdot{r2,l2}
\end{fmfgraph*}
\end{fmffile}
\end{center}
}=\frac{i \bar{f}(v_\alpha k/T)}{\omega + v_\alpha k + i0^+}.
\end{eqnarray}
The one-loop contribution to the self-energy $\Sigma_n$ can thus be computed. The real part of it is just a renormalization of $E_n$, while the imaginary part, important when $\omega\simeq E_n$, is half of the width, $\Gamma_n/2$; it is given by $U^2$ times the phonon density of states at energy $|E_n-E_m|$, where $U$ is the energy scale of the interaction vertex and $E_m$ is the energy of the intermediate defect states.

Now we show that higher loop diagrams give corrections to $\Gamma$ that are higher order in $U/\epsilon$. This conclusion is not as obvious as it might seem. Suppose there are $l$ phonon loops in a self-energy diagram. While there will be $2l$ factors of $U$, which is small, there are also $2l-1$ defect Green's functions that may each be at resonance, which can potentially be large. More particularly, suppose all contributions up to $(l-1)$-loop to the defect self-energy has been computed, giving a width $\Gamma_{(l-1)}$ (and we already know $\Gamma_{(1)}$ is order $U^2$), then, when we consider the correction from $\Gamma_{(l-1)}$ to $\Gamma_{(l)}$ by $l$-loop diagrams, a naive estimation suggests that, when the $2l-1$ defect Green's functions (whose widths are self-consistently taken to be $\Gamma_{(l-1)}$) are all at resonance, they contribute a factor of $\Gamma_{(l-1)}^{-(2l-1)}$, and this requires all the $l$ internal loop phonons to be within a width of $\Gamma_{(l-1)}$, giving rise to a factor of $\Gamma_{(l-1)}^l$. Combined with the $U^{2l}$ factor from the vertices, we have
\begin{align}
\mbox{naive correction from $\Gamma_{(l-1)}$ to $\Gamma_{(l)}$:} \ \ \ \ U^{2l} \Gamma_{(l-1)}^{1-l} \ .
\end{align}
By induction, this would seem to suggest that up to any loop number $l$, the self-consistent contribution $\Gamma_{(l)}$ to $\Gamma$ is always of order $U^2$, which would lead to a problem, since they are all as large as the 1-loop $\Gamma_{(1)}$. Fortunately, a closer inspection shows this is not the case. In a higher loop diagram, there is at least one phonon whose on-shell energy $\omega=vk$ will appear in $r>1$ many defect Green's functions with the same sign. Suppose $r=2$. In the loop integral of this phonon, we will have a factor
\begin{align}
\int_0^{\omega_{c}} d\omega \ F(\omega) \ \frac{1}{\omega-Z_2 + i\Gamma} \frac{1}{\omega-Z_1 + i\Gamma}
\end{align}
(or with $-\omega$ in place of $\omega$ in the denominators, depending on the arrow direction of the phonon) where $F(\omega)$ is some regular function of $\omega$ that vanishes as $\omega\rightarrow 0$, and $Z_1, Z_2$ are some energies that depend on the energies of the intermediate defect states and other intermediate phonon states. When $|Z_1-Z_2|> \Gamma$ the worried simultaneous resonance does not occur. When it does occur, $Z_1\sim Z_2\sim Z$, we express the $\omega$ integral as
\begin{align}
\int_0^{\omega_{c}} d\omega = \left( \int_{-\infty}^{\infty} - \int_{\omega_c}^{\infty}  - \int_{-\infty}^{0} \right) d\omega \
\end{align}
($F(\omega)$ should be defined only between $0$ and $\omega_c$; in order for this rewriting of the integral to make sense, we can regularize $F(\omega)$ in any way such that it vanishes as $\omega\rightarrow\pm \infty$). For the first term we can use contour integral, giving rise to $\sim \partial_Z F(Z)$, which is much smaller than the large factor $F(Z)\Gamma/\Gamma^2$ that we would have worried about. For the second term, since $Z\ll \omega_c$ by assumption, there is no contribution to the imaginary part from resonance. For the last term, if $|Z| \gg \Gamma$ there is again no contribution the imaginary part; when $|Z|$ is close $0$ the situation is more delicate: we need to resort to the fact that, for phonons with small on-shell energy $\omega$, the density of phonon states as well as the interaction vertices that are included in $F(\omega)$ always make $F(\omega)$ scale with a power higher than $\omega^2$, hence again there is no large factor (if we were not dealing with phonons but photons -- the crucial difference being Goldstone boson versus gauge boson -- there would be less derivatives in the interaction vertices, making $F(\omega)$ scale with lower power, in which case a more thorough analysis is needed, which is the famous resolution to the infra-red ``divergence'' problem). For larger $r$, similar reasoning apply. Therefore, we can conclude that even at resonance, at leading order in self-consistent perturbation theory it is sufficient to consider 1-loop self-energy.

The validity of the second part of the recipe, that we only need to consider the one phonon in and one phonon out resonant scattering, now follows. Suppose the defect state $|i\rangle$ absorbs a phonon and is now at a resonant intermediate state $|m\rangle$. The intermediate state $|m\rangle$ may in general decay to some final state $|f\rangle$ by emitting some phonons into and absorbing some phonons from the thermal ensemble. By the optical theorem, the probability of having a totally number of $l$ phonons being emitted and absorbed is given by the previously estimated $\Gamma_{(l)}-\Gamma_{(l-1)}$, and since only $\Gamma_{(1)}$ is at leading order in self-consistent perturbation theory, we arrive at our claim.

Before we close, we make two further remarks. First, suppose there are two resonant processes, one with energy difference $\epsilon \sim T$ and the other with energy difference $\epsilon'\ll\epsilon$, both contributing to a same physical quantity of interest. Then the contribution of the second process is in general unimportant compared to the first one. This is because while the Boltzmann weight for phonons involved in the second process is larger by at factor of $\sim T/\epsilon'$ (compared to those involved the first process, same below), these phonon' density of states as well as their accompanied derivatives in the interaction vertices introduce a suppression by a higher power in $\epsilon'/\epsilon$. This is why we have ignored the transition between $|\pm\rangle$ in our three-level model. 

The second remark is, a resonant intermediate state is counted as intermediate during the scattering process only if the lifetime $\sim 1/\Gamma$ is shorter than the total time available for the scattering process. In the context of scattering off disorders, the available time should be the typical relaxation time $\tau$. This is generally true for our interested systems.

\end{widetext}

\bibliography{phonon}{}

\bibliographystyle{apsrev4-2}  

\end{document}